# Which Clustering Do You Want?
# Inducing Your Ideal Clustering with Minimal Feedback

**Sajib Dasgupta**                                    SAJIB@HLT.UTDALLAS.EDU
**Vincent Ng**                                        VINCE@HLT.UTDALLAS.EDU
*Human Language Technology Research Institute*
*University of Texas at Dallas*
*800 West Campbell Road; Mail Station EC31*
*Richardson, TX 75080-3021 U.S.A.*

## Abstract

While traditional research on text clustering has largely focused on grouping documents by topic, it is conceivable that a user may want to cluster documents along other dimensions, such as the author's mood, gender, age, or sentiment. Without knowing the user's intention, a clustering algorithm will only group documents along the most prominent dimension, which may not be the one the user desires. To address the problem of clustering documents along the user-desired dimension, previous work has focused on learning a similarity metric from data manually annotated with the user's intention or having a human construct a feature space in an interactive manner during the clustering process. With the goal of reducing reliance on human knowledge for fine-tuning the similarity function or selecting the relevant features required by these approaches, we propose a novel active clustering algorithm, which allows a user to easily select the dimension along which she wants to cluster the documents by inspecting only a small number of words. We demonstrate the viability of our algorithm on a variety of commonly-used sentiment datasets.

## 1. Introduction

Text clustering is one of the major application domains for demonstrating the viability of a clustering algorithm. While traditional research on text clustering has largely focused on grouping documents by topic, it is conceivable that a user may want to cluster documents along other *dimensions*, such as the author's mood, gender, age, or sentiment. Since virtually all existing text clustering algorithms can produce just *one* clustering of a given set of documents, a natural question is: is this clustering necessarily the one the user desires? In other words, can a text clustering algorithm always produce a clustering along the user-desired dimension?

The answer to this question depends to a large extent on whether the user can successfully "communicate" her intention to the clustering algorithm. Traditionally, this can be achieved by designing a good similarity function that can capture the similarity between a pair of documents, so that her *ideal* clustering can be produced. This typically involves having her identify a set of features that is useful for inducing the desired clusters (Liu, Li, Lee, & Yu, 2004). However, manually identifying the "right" set of features is both time-consuming and knowledge-intensive, and may even require a lot of domain expertise. The fact that the resulting similarity function is typically not easily portable to other domains is particularly unappealing from a machine-learning perspective. To overcome this weakness, researchers have attempted to *learn* a similarity metric from *side* information (Xing, Ng, Jordan, & Russell, 2002), such as constraints on which pairs of documents must or must not appear in the same cluster (Wagstaff, Cardie, Rogers, & Schrödl, 2001).

By contrast, recent work has focused on *active clustering*, where a clustering algorithm can incorporate user feedback *during* the clustering process to help ensure that the documents are grouped according to the user-desired dimension. One way to do this is to have the user incrementally construct a set of relevant features in an interactive fashion (Bekkerman, Raghavan, Allan, & Eguchi, 2007; Raghavan & Allan, 2007; Roth & Small, 2009). Another way is to have the user correct the mistakes made by the clustering algorithm in each clustering iteration by specifying whether two existing clusters should be *merged* or *split* (Balcan & Blum, 2008). A major drawback associated with these active clustering algorithms is that they involve a considerable amount of human feedback, which needs to be provided in *each* iteration of the clustering process. Furthermore, identifying clusters for merging or splitting in Balcan and Blum's algorithm may not be as easy as it appears: for each MERGE or SPLIT decision the user makes, she has to sample a large number of documents from the cluster(s), read through the documents, and base her decision on the extent to which the documents are (dis)similar to each other.

In this article, we attack the problem of clustering documents according to user interest from a different angle. We aim to have a *knowledge-lean* approach to this problem — an approach that can produce a clustering of the documents along the user-desired dimension without relying on human knowledge for fine-tuning the similarity function or selecting the relevant features, unlike existing approaches. To this end, we propose a novel active clustering algorithm, which assumes as input a simple feature representation (composed of unigrams only) and a simple similarity function (i.e., the dot product), and operates by (1) inducing the important *clustering dimensions*[1] of a given set of documents, where each clustering dimension is represented by a (small) number of automatically selected words that are representative of the dimension; and (2) have the user choose the dimension along which she wants to cluster the documents by examining these automatically selected words. In comparison to the aforementioned feedback mechanisms, ours is arguably much simpler: we only require that the user have a cursory look at a small number of features for each dimension once and for all, as opposed to having the user *generate* the feature space in an interactive manner or *identify* clusters that need to be merged or split in *each* clustering iteration.

We evaluate our active clustering algorithm on the task of *sentiment-based* clustering, where the goal is to cluster a set of documents (e.g., reviews) according to the polarity (e.g., "thumbs up" or "thumbs down") expressed by the author without using any labeled data. Our decision to focus on sentiment-based clustering is motivated by several reasons. One reason is that there has been relatively little work on sentiment-based clustering. As mentioned before, existing work on text clustering has focused on *topic-based* clustering, where high accuracies can be achieved even for datasets with a large number of classes (e.g., 20 Newsgroups); and despite the large amount of recent work on sentiment analysis and opinion mining, much of it has focused on *supervised* methods (see Pang & Lee, 2008, for a comprehensive survey of the field).

---

1. We use the term *clustering dimension* to refer to a dimension along which a set of documents can be clustered. For example, a set of movie reviews can be clustered according to genre (e.g., action, romantic, or documentary) or sentiment (e.g., positive, negative, neutral).

| Review 1 |
|---|
| The sound from my system did seem to be a little better |
| (the CD's were not skipping as much). But the bottom line is it |
| didn't fix the problem as the CDs are still skipping noticeably, |
| although not as bad as before. ... |
| Review 2 |
| John Lynch wrote a classic in Spanish-American Revolutions 1808-1826. |
| He describes all the events that led to the independence of Latin America from Spain. |
| The book starts in Rio de La Plata and ends in Mexico and Central America. |
| Curiously one can note a common pattern of highly stratified societies lead by Spanish ... |
| The reluctance of Spanish Monarchy (and later even of liberals) led to independence ... |
| For all of those who are interested in a better understanding of Latin ??this great book is a must. |
| Lynch cleverly combines historical and economic facts about the Hispanic American societies ... |

Table 1: Snippets of two reviews that illustrate the two challenges of polarity classification. One is that reviews are sentimentally ambiguous (Review 1), and the other is that the objective materials in a review can significantly outnumber their subjective counterparts (Review 2).

Another equally important reason for our focus on sentiment-based clustering is concerned with the challenges that this task presents to natural language processing (NLP) researchers. Broadly speaking, the complexity of sentiment-based clustering arises from two sources. First, reviews are *sentimentally ambiguous*, containing both positive and negative sentiment-bearing words and phrases. Review 1 of Table 1 shows a snippet of a review from the DVD domain that illustrates the sentimental ambiguity problem: while the phrases "a little better", "not skipping", and "not as bad" convey a positive sentiment, the phrases "didn't fix" and "skipping noticeably" are negative sentiment-bearing. Hence, unless a sentiment analyzer performs deeper linguistic analysis, it will be difficult for the analyzer to determine the polarity of the review. Second, the objective materials in a review tend to significantly outnumber their subjective counterparts, as a reviewer typically devotes a large portion of the review to describing the features of a product before assigning a rating to it; consequently, any sentiment analyzer that uses a word- or phrase-based feature representation will be composed of mostly features that are irrelevant with respect to polarity determination. Shown in Review 2 of Table 1 is a snippet of a book review that illustrates this problem. As we can see, all but three words/phrases ("classic", "great book", "cleverly") in this review correspond to objective materials.

The aforementioned complications present significant challenges even to *supervised* polarity classification systems, let alone sentiment-based clustering algorithms, which do not have access to any labeled data. To further illustrate the difficulty that these two complications impose on sentiment-based clustering, consider the task of clustering a set of movie reviews. Since each review may contain a description of the plot and the author's sentiment, a clustering algorithm may cluster reviews along either the *plot* dimension or the *sentiment* dimension; and without knowing the user's intention, they will be clustered along the most *prominent* dimension. Assuming the usual bag-of-words representation, the most prominent dimension will more likely be *plot*, as it is not uncommon for a review to be devoted almost exclusively to the plot, with the author briefly expressing her sentiment only at the end of the review. Even if the reviews contain mostly subjective materials, the most prominent dimension may still not be *sentiment* owing to the aforementioned sentimental ambiguity problem: the presence of both positive and negative sentiment-bearing words in these reviews renders the sentiment dimension *hidden* (i.e., less prominent) as far as clustering is concerned.

In sum, our contributions in this article are five-fold.

- We propose a novel active clustering algorithm that can cluster a set of documents along the user-desired dimension without any labeled data or side information such as manually specified or automatically acquired must-link and cannot-link constraints. In comparison to existing active clustering approaches, our algorithm has the appeal of requiring much simpler human feedback.

- We demonstrate the viability of our algorithm not only by evaluating its performance on sentiment datasets, but also via a set of human experiments, which is typically absent in papers that involve algorithms for incorporating user feedback.

- Our results have led to a deeper understanding of spectral clustering. Specifically, we propose a novel application of the top eigenvectors produced by a spectral clustering algorithm, where we use them to unveil the important clustering dimensions of a text collection.

- Our results also have implications for domain adaptation, a topic that has recently received a lot of attention in the NLP community. Specifically, we show that the sentiment dimension manually identified for one domain can be used to automatically identify the sentiment dimension for a new, but similar, domain.

- Preliminary results on datasets that possess more than one clustering dimension (e.g., a collection of book and DVD reviews, which can be clustered by sentiment or by the type of the product concerned) indicate that our algorithm is capable of producing *multiple* clusterings of a dataset, one along each dimension. Hence, our algorithm can potentially reveal more information from a dataset than is possible with traditional clustering algorithms, which can only produce a single clustering of the data. The ability to produce multiple clusterings is a particularly useful feature for a user who does not have any idea of how she wants the documents to be clustered (due to the lack of knowledge of the data, for instance). Even if a user has some knowledge of the data and knows how she wants the documents to be clustered, our algorithm can help unveil other "hidden" dimensions that she is not previously aware of but may also be of interest to her.

The rest of this article is organized as follows. Section 2 presents the basics of spectral clustering, which will facilitate the discussion of our active clustering algorithm in Section 3. We describe our human experiments and evaluation results on several sentiment datasets in Section 4 and the significance of our work in Section 5. Finally, we discuss related work in Section 6 and conclude in Section 7.

## 2. Spectral Clustering

When given a clustering task, an important question to ask is: which clustering algorithm should we use? A popular choice is $k$-means. Nevertheless, it is well-known that $k$-means has the major drawback of not being able to separate data points that are not linearly separable in the given feature space (e.g., see Dhillon, Guan, & Kulis, 2004; Cai, He, & Han, 2005). Moreover, since $k$-means clusters documents directly in the given feature space, which for text applications typically comprises hundreds of thousands of features, its performance could be adversely affected by the *curse of dimensionality*. Spectral clustering algorithms were developed in response to these problems with $k$-means. In this section, we first present one of the most commonly-used algorithms for spectral clustering (Section 2.1). Then, we provide the intuition behind spectral clustering (Section 2.2). Finally, we describe two ways to use the resulting eigenvectors to produce a clustering (Section 2.3).

### 2.1 Algorithm

Let $X=\{x_1,\ldots,x_n\}$ be a set of $n$ data points to be clustered, $s : X \times X \to \Re$ be a similarity function defined over $X$, and $S$ be a similarity matrix that captures pairwise similarities (i.e., $S_{i,j} = s(x_i, x_j)$). Like many other clustering algorithms, a spectral clustering algorithm takes $S$ as input and outputs a $k$-way partition $C = \{C_1, C_2, .., C_k\}$ (i.e., $\cup_{i=1}^k C_i = X$ and $\forall i, j : i \neq j \implies C_i \cap C_j = \emptyset$). Equivalently, one can think of spectral clustering as learning a partitioning function $f$, which, in the rest of this article, will be represented as a vector such that $f(i) \in \{1, \ldots, k\}$ indicates the cluster to which $x_i$ should be assigned. Note that the cluster labels are interchangeable and can even be renamed without any loss of generality.

Among the well-known spectral clustering algorithms (e.g., Weiss, 1999; Shi & Malik, 2000; Kannan, Vempala, & Vetta, 2004), we adopt the one proposed by Ng, Jordan, and Weiss (2001), as it is arguably the most widely-used. Below are the main steps of Ng et al.'s spectral clustering algorithm:

1. Create the diagonal matrix $D$ whose $(i,i)$-th entry is the sum of the $i$-th row of $S$, and then construct the Laplacian matrix[2] $L = D^{-1/2} S D^{-1/2}$.

2. Find the eigenvalues and the eigenvectors of $L$.

3. Create a new matrix from the $m$ eigenvectors that correspond to the $m$ largest eigenvalues.

4. Each data point is now rank-reduced to a point in the $m$-dimensional space. Normalize each point to unit length (while retaining the sign of each value).

5. Apply $k$-means to cluster the data points using the resulting $m$ eigenvectors.

In other words, spectral clustering clusters data points in a low-dimensional space, where each dimension corresponds to a top eigenvector of the Laplacian matrix.

---

2. We follow Ng et al. (2001) and employ a normalized dual form of the usual Laplacian $D - S$.

## 2.2 Intuition behind Spectral Clustering

It may not be immediately clear why spectral clustering produces a "meaningful" partitioning of a set of points. There are theoretical justifications behind spectral clustering, but since the mathematics is quite involved, we will only provide an intuitive justification of this clustering technique in a way that is sufficient for the reader to understand our active clustering algorithm in Section 3, and refer the interested reader to Shi and Malik's (2000) seminal paper on spectral clustering for details. Since we will only apply spectral clustering to produce a 2-way clustering of a given set of data points in the rest of this article, we will center our discussion on 2-way clustering in this subsection.

Spectral clustering employs a graph-theoretic notion of grouping. Specifically, a set of data points in an arbitrary feature space is represented as an undirected weighted graph, where each node corresponds to a data point, and the edge weight between two nodes $x_i$ and $x_j$ is their similarity, $S_{i,j}$.

Given this graph formulation, a reasonable way to produce a 2-way partitioning of the data points is to minimize the similarity between the resulting two clusters, $C_1$ and $C_2$. Hence, a reasonable objective function to minimize is the *cut value*, where

$$Cut(C_1, C_2) = \sum_{i,j} S_{i,j}(f(i) - f(j))^2.$$

Without loss of generality, we can define $f$ as follows.

$$f(i) = \left\{ \begin{array}{ccc} 1 & : & i \in C_1 \\ -1 & : & i \in C_2 \end{array} \right.$$

As mentioned before, while we use 1 and $-1$ as cluster labels here, they are interchangeable and can in fact be renamed in whatever way we want.

One problem with minimizing the cut value, as noticed by Wu and Leahy (1993), is that this objective favors producing unbalanced clusters in which one of them contains a very small number of nodes. In other words, there is a bias towards isolating a small set of nodes. As mentioned by Shi and Malik (2000), this should not be surprising, since the number of edges involved in the cut (and hence the cut value) tends to increase as the sizes of the two clusters become relatively balanced.

A closer examination of the minimum cut criterion reveals the problem: while it minimizes inter-cluster similarity, it makes no attempt to maximize intra-cluster similarity. To address this weakness, Shi and Malik (2000) propose to minimize instead the *normalized* cut value, $NCut$, which takes into account both inter-cluster dissimilarity and intra-cluster similarity. More specifically,

$$NCut(C_1, C_2) = \frac{Cut(C_1, C_2)}{assoc(C_1, C_1 \cup C_2)} + \frac{Cut(C_1, C_2)}{assoc(C_2, C_1 \cup C_2)},$$

where $assoc(A, B)$, computed as $\sum_{x_i \in A, x_j \in B} S_{i,j}$, is the total connection from the nodes in $A$ to the nodes in $B$. Given this definition, a cut resulting from unbalanced clusters will no longer have a small $NCut$ value. To see the reason, consider the case where $C_1$ consists of just one node. In this case, $assoc(C_1, C_1 \cup C_2) = Cut(C_1, C_2)$, making $NCut(C_1, C_2)$ large.

After some algebra, we can express $NCut$ as follows:

$$NCut(C_1, C_2) = \frac{f^T(D-S)f}{f^T D f}$$

subject to the constraints that $(Df)^T \mathbf{1} = 0$ and

$$f(i) = \begin{cases} \sqrt{\frac{\sum_{i \in C_2} d(i)}{\sum_{i \in C_1} d(i)}} & : \quad i \in C_1 \\ -\sqrt{\frac{\sum_{i \in C_1} d(i)}{\sum_{i \in C_2} d(i)}} & : \quad i \in C_2 \end{cases}$$

where $d(i) = D(i,i)$, as defined in Section 2.1.[3] The first constraint, which specifies that $Df$ is orthogonal to $\mathbf{1}$, can be intuitively understood as follows: since $\mathbf{1}$, being a *constant* vector where all of its entries are 1, cannot be used to induce a partition, this constraint avoids the trivial solution in which all points are assigned to the same cluster.

Unfortunately, Papadimitriou proves that minimizing normalized cut is an NP-complete problem, even for the special case of graphs on regular grids (see Shi & Malik, 2000, for the proof). Hence, following Shi and Malik, we relax this minimization problem by dropping the second constraint and allowing each entry of $f$ to take on a real value rather than one of two discrete values, seeking a real-valued solution to the following problem:

$$\min_{f \in \Re^n} \frac{f^T(D-S)f}{f^T D f} \tag{1}$$

subject to

$$Df \perp \mathbf{1}.$$

Assuming that $g = D^{1/2}f$, we can rewrite Problem (1) as

$$\min_{g \in \Re^n} \frac{g^T D^{-1/2}(D-S)D^{-1/2}g}{g^T g} \tag{2}$$

subject to

$$g \perp D^{1/2}\mathbf{1}.$$

Following the standard Rayleigh-Ritz theorem, one can prove that the solution to Problem (2), $g$, is the eigenvector that corresponds to the second smallest eigenvalue of $D^{-1/2}(D-S)D^{-1/2}$, or equivalently, the eigenvector that corresponds to the second largest eigenvector of $D^{-1/2}SD^{-1/2}$, which is the Laplacian matrix $L$ defined in Section 2.1. For simplicity, we will henceforth refer to the eigenvector that corresponds to the $n$-th largest eigenvalue of $L$ simply as its $n$-th eigenvector and denote it as $\mathbf{e}_n$.[4]

---

3. Besides normalized cut, ratio cut (Chan, Schlag, & Zien, 1994), average association (Shi & Malik, 2000), and min-max cut (Ding, He, Zha, Gu, & Simon, 2001) have also been used as objective functions for spectral clustering algorithms.

4. Given that Problem (2) involves minimizing a Rayleigh quotient, it may seem somewhat unintuitive that its solution is the second eigenvector of $L$ rather than its first eigenvector. The reason can be attributed to the constraint associated with the problem, which specifies that the solution $g$ is perpendicular to $D^{1/2}\mathbf{1}$, the first eigenvector of $L$.

This is the idea behind spectral clustering: the second eigenvector of $L$ is an *approximate* solution to the problem of minimizing normalized cut.[5] Of course, since the second eigenvector is a real-valued solution, we will have to convert it into a partitioning function so that it can be used to cluster the data points. Section 2.3 explains two simple ways of converting this eigenvector into a partitioning function.

It turns out that the other eigenvectors of $L$ also convey useful information about the data. Specifically, if we impose an additional constraint to Problem (2) forcing the solution to be orthogonal to the second eigenvector of $L$, then the solution becomes the third eigenvector. Hence, the third eigenvector can be thought of as a suboptimal solution to Problem (2), meaning that it can also be used to impose a "reasonably good" partition of the data points. Perhaps more importantly, since the eigenvectors of $L$ are orthogonal to each other (because $L$ is symmetric), the clustering produced by using the third eigenvector is likely to correspond to a different dimension of the data than that produced by the second eigenvector.

More generally, if we limit the solution space to only those real-valued vectors that are orthogonal to the first $m$ eigenvectors of $L$, then the solution to our constrained optimization problem is the $(m + 1)$-th eigenvector of $L$. In other words, each of the top eigenvectors of $L$ can intuitively be thought of as revealing an important dimension of the data, although subsequent eigenvectors are progressively less ideal as far as clustering is concerned.

## 2.3 Clustering with Eigenvectors

As Ng et al. (2001) point out, "different authors still disagree on which eigenvectors to use, and how to derive clusters from them". In this subsection, we describe two common methods for determining which eigenvectors to use, and for each method, we show how to derive clusters using the selected eigenvector(s). These methods will serve as baselines in our evaluation.

### 2.3.1 METHOD 1: USING THE SECOND EIGENVECTOR ONLY

Since Shi and Malik (2000) show that the second eigenvector, $\mathbf{e}_2$, is the approximate solution to the problem of minimizing the normalized cut, it should perhaps not be surprising that $\mathbf{e}_2$ is commonly chosen as the only eigenvector for deriving a partition. However, since $\mathbf{e}_2$ is a real-valued solution to the constrained optimization problem, we need to specify how we can derive clusters from it.

Clustering using $\mathbf{e}_2$ is trivial: since we have a linearization of the points, one simple way is to determine the threshold for partitioning them. However, we follow Ng et al. (2001) and cluster the points using 2-means in this one-dimensional space.

### 2.3.2 METHOD 2: USING THE TOP $m$ EIGENVECTORS

Recall from Section 2.1 that after eigen-decomposing the Laplacian matrix, each data point is represented by $m$ co-ordinates. In the second method, we use 2-means to cluster the data points in this $m$-dimensional space, effectively exploiting all of the top $m$ eigenvectors.

---

5. In fact, since $f = D^{-1/2}g$, we have to pre-multiply the second eigenvector of $L$ by $D^{-1/2}$ to get the solution to Problem (1), but following Ng et al. (2001), we employ the second eigenvector of $L$ directly for clustering, ignoring the term $D^{-1/2}$.

## 3. Our Active Clustering Algorithm

As mentioned before, sentiment-based clustering is challenging, in part due to the fact that reviews can be clustered along more than one dimension. In this section, we describe our active clustering algorithm, which makes it easy for a user to specify that the dimension along which she wants to cluster the data points is sentiment. Recall that our algorithm first applies spectral clustering to reveal the most important dimensions of the data, and then lets the user select the desired dimension (i.e., sentiment). To motivate the importance of user feedback, it helps to understand why the two baseline clustering algorithms described in Section 2.3, which are also based on spectral methods but do not rely on user feedback, may not always yield a sentiment-based clustering. To begin with, consider the first method, where only the second eigenvector is used to induce the partition. Recall that the second eigenvector reveals the most prominent dimension of the data. Hence, if sentiment is not the most prominent dimension (which can happen if the non-sentiment-bearing words outnumber the sentiment-bearing words in the bag-of-words representation of a review), then the resulting clustering of the reviews may not be sentiment-oriented. A similar line of reasoning can be used to explain why the second baseline clustering algorithm, which clusters based on all of the top eigenvectors, may not always work well. Since each eigenvector corresponds to a different dimension (and, in particular, some of them correspond to non-sentiment dimensions), using all of them to represent a review may hamper the accurate computation of the similarity of two reviews as far as clustering along the sentiment dimension is concerned. In the rest of this section, we discuss in detail the major steps of our active clustering algorithm, which allows easy incorporation of user feedback.

### 3.1 Step 1: Identify the Important Clustering Dimensions

We rely on a simple method for identifying the important clustering dimensions of a given text collection: we employ the top eigenvectors of the Laplacian as the important clustering dimensions. This method is motivated by the fact that $\mathbf{e}_2$, the second eigenvector of the Laplacian, is the optimal real-valued solution to the objective function that spectral clustering minimizes (i.e., normalized cut, Shi & Malik, 2000), and is therefore an optimal clustering dimension. More importantly, we exploit a rarely-utilized observation discussed in Section 2.2: while the remaining eigenvectors are all suboptimal solutions (with $\mathbf{e}_i$ being more suboptimal as $i$ increases), the top eigenvectors (i.e., those with small $i$ values), being less suboptimal, may still yield reasonably good (though not optimal) clusterings of the data and can therefore serve as good clustering dimensions. Existing applications of spectral clustering have mainly clustered data points in the space defined by *all* of the top eigenvectors, and have not attempted to use each of the $\mathbf{e}_i$s (with $i > 2$) separately to produce clusterings, unlike ours. Note that the first eigenvector, being a constant vector, simply assigns all data points to the same cluster and therefore is typically ignored.

### 3.2 Step 2: Identify the Relevant Features for Each Partition

Given the eigen-decomposition from Step 1, we first obtain the second through the $m$-th eigenvectors, which correspond to the most important dimensions of the data. The next question is: how can we determine which dimension captures the user interest? One way to do this is to have the user inspect each of the $m-1$ partitions of the reviews and decide which corresponds most closely to a sentiment-based clustering. The main drawback associated with this kind of user feedback is that the user may have to read a large number of reviews in order to make a decision. Hence, to reduce human effort, we employ an alternative procedure: we (1) identify the most informative features for characterizing each partition, and (2) have the user inspect just the features rather than the reviews. To make it easy for a human to identify a clustering dimension, the features should be chosen so that they are useful for distinguishing the reviews in the two clusters.

To identify and rank the informative features, we employ a method that we call *maximum margin feature ranking* (MMFR).[6] Recall that a maximum margin classifier (e.g., a support vector machine) separates data points from two classes while maximizing the margin of separation. Specifically, a maximum margin hyperplane is defined by $\mathbf{w} \cdot \mathbf{x} - b = 0$, where $\mathbf{x}$ is a feature vector representing an arbitrary data point, and $\mathbf{w}$ (a weight vector) and $b$ (a scalar) are parameters that are learned by solving the following constrained optimization problem:

$$\min \ \frac{1}{2}\|\mathbf{w}\|^2 + C \sum_i \xi_i$$

subject to

$$c_i(\mathbf{w} \cdot \mathbf{x}_i - b) \geq 1 - \xi_i, \quad 1 \leq i \leq n,$$

where $c_i \in \{+1, -1\}$ is the class of the $i$-th training point $\mathbf{x}_i$, $\xi_i$ is the degree of misclassification of $\mathbf{x}_i$, and $C$ is a regularization parameter that balances training error and model complexity.

We use $\mathbf{w}$ to identify the most informative features for a partition. Note that the most informative features are those with large absolute weight values: a feature with a large positive (negative) weight is strongly indicative of the positive (negative) class.[7] We exploit this observation and identify the most informative features for a partition by (1) training a binary SVM classifier[8] on the partition, where data points in the same cluster are assumed to have the same class value; (2) sorting the features according to the SVM-learned feature weights; and (3) generating two ranked lists of informative features using the top and bottom $F$ features, respectively.

Given the ranked lists generated for each of the $m-1$ partitions, the user will select one of the partitions/dimensions as most relevant to sentiment by inspecting as many features in the ranked lists as needed. After picking the most relevant dimension, the user will label one of the two feature lists associated with this dimension as POSITIVE and the other as NEGATIVE. Since each feature list represents one of the clusters, the cluster associated with the positive list is labeled POSITIVE and the cluster associated with the negative list is labeled NEGATIVE.

---

6. Note that other commonly-used feature selection techniques such as log-likelihood ratio and information gain can also be applied to identify these informative features (see Yang & Pedersen, 1997, for an overview).

7. The notion of using SVM feature weights as measures of feature informativeness has also been explored in other work. See, for instance, the work of Fung (2003), Gilad-Bachrach, Navot, and Tishby (2004), and Kugler, Aoki, Kuroyanagi, Iwata, and Nugroho (2005) for details.

8. All the SVM classifiers in this article are trained using the SVM$^{light}$ package (Joachims, 1999a), with the learning parameters set to their default values.

In comparison to existing user feedback mechanisms for assisting a clustering algorithm, ours requires comparatively little human intervention: we only require that the user select a dimension by examining a small number of features, as opposed to having the user construct the feature space or identify clusters that need to be merged or split as is required with other methods.

### 3.3 Step 3: Identify the Unambiguous Reviews

There is a caveat, however. As mentioned in the introduction, many reviews contain both positive and negative sentiment-bearing words. These ambiguous reviews are more likely to be clustered incorrectly than their unambiguous counterparts. Since the ranked lists of features are derived from each partition, the presence of these ambiguous reviews can adversely affect the identification of informative features using MMFR. As a result, we remove the ambiguous reviews before deriving informative features from a partition.

We employ a simple method for identifying unambiguous reviews. In the computation of eigenvalues, each data point factors out the orthogonal projections of each of the other data points with which they have an affinity. Ambiguous data points receive the orthogonal projections from both the positive and negative data points, and hence they have near zero values in the pivot eigenvectors. In other words, the points with near zero values in the eigenvectors are more ambiguous than those with large absolute values. We therefore sort the data points according to their corresponding values in the eigenvector, and keep only the top $n/8$ and the bottom $n/8$ data points. We induce the informative features only from the resulting 25% of the data points, and present them to the user so that she can select the desired partition.[9]

### 3.4 Step 4: Cluster Along the Selected Eigenvector

Finally, we employ 2-means to cluster all the reviews along the eigenvector selected by the user, regardless of whether a review is ambiguous or not.

## 4. Evaluation

In this section, we describe experiments that aim to evaluate the effectiveness of our active clustering algorithm and provide insights into it.

### 4.1 Experimental Setup

We begin by discussing the details on the datasets, the document preprocessing method, the implementation of spectral clustering, and the evaluation metrics.

---

9. Note that 25% is a somewhat arbitrary choice. Underlying this choice is merely the assumption that a fraction of the reviews is unambiguous. As we will see in the evaluation section, these reviews can be classified according to their polarity with a high accuracy; consequently, the features induced from the resulting clusters are also of high quality. Additional experiments revealed that the list of top-ranking features does not change significantly when induced from a smaller number of unambiguous reviews.

### 4.1.1 Datasets

We use five sentiment datasets, including the widely-used movie review dataset [MOV] (Pang, Lee, & Vaithyanathan, 2002) as well as four datasets containing reviews of four different types of products from Amazon [Books (BOO), DVDs (DVD), Electronics (ELE), and Kitchen Appliances (KIT)] (Blitzer, Dredze, & Pereira, 2007). Each dataset has 2000 labeled reviews (1000 positives and 1000 negatives). To illustrate the difference between topic-based clustering and sentiment-based clustering, we will also show topic-based clustering results on POL, a dataset created by taking all the documents from the two sections of 20 Newsgroups that discuss issues in cryptography and politics, namely, `sci.crypt` and `talks.politics`.

### 4.1.2 Document Preprocessing

To preprocess a document, we first tokenize and downcase it, and then represent it as a vector of unstemmed unigrams, each of which assumes a value of 1 or 0 that indicates its presence or absence in the document. In addition, we remove from the vector punctuation, numbers, words of length one, and words that occur in only a single review.

Following common practice in the information retrieval community, we also exclude words with a high document frequency, many of which are stopwords or domain-specific general-purpose words (e.g., "movies" in the movie domain). A preliminary examination of our evaluation datasets reveals that these words typically comprise 1–2% of a vocabulary. The decision of exactly how many terms to remove from each dataset is subjective: a large corpus typically requires more removals than a small corpus. To be consistent, we simply sort the vocabulary by document frequency and remove the top 1.5%. We will henceforth refer to this document representation as the *bag-of-words* (BOW) representation.

### 4.1.3 Spectral Learning Setup

Following common practice in spectral learning for text domains (e.g., Kamvar, Klein, & Manning, 2003; Cai et al., 2005), we compute the similarity between two reviews by taking the dot product of their feature vectors. As in Ng et al.'s (2001) spectral clustering algorithm, we set the diagonal entries of the similarity matrix to 0. In addition, we set $m$ to 5. In other words, we consider the second through fifth eigenvectors, assuming that they are sufficient for capturing the desired clusterings.[10]

### 4.1.4 Evaluation Metrics

We employ two evaluation metrics. First, we report results for each dataset in terms of accuracy, which is the percentage of documents for which the label assigned by our system is the same as the gold-standard label. Second, following Kamvar et al. (2003), we evaluate the clusters produced by our approach against the gold-standard clusters using the Adjusted Rand Index (ARI), which is the corrected-for-chance version of the Rand Index. More specifically, given a set of $N$ data points and two clusterings of these points, $U$ and $V$,

---

10. Note that setting $m$ to 5 is a somewhat arbitrary choice, and that any number of eigenvectors can be used in our active clustering algorithm.

where $U = \{U_1, U_2, \ldots, U_m\}$ has $m$ clusters and $V = \{V_1, V_2, \ldots, V_n\}$ has $n$ clusters, ARI is computed as follows:

$$ARI(U, V) = \frac{\sum_{ij} \binom{n_{ij}}{2} - [\sum_i \binom{a_i}{2} \sum_j \binom{b_j}{2}]/\binom{N}{2}}{\frac{1}{2}[\sum_i \binom{a_i}{2} + \sum_j \binom{b_j}{2}] - [\sum_i \binom{a_i}{2} \sum_j \binom{b_j}{2}]/\binom{N}{2}}$$

In this formula, $n_{ij}$ is the number of common objects in $U_i$ and $V_j$; whereas $a_i$ and $b_j$ are the number of objects in $U_i$ and $V_j$, respectively. ARI ranges from $-1$ to $1$; better clusterings have higher ARI values.

## 4.2 Baseline Systems

In this subsection, we describe our baseline results. The first two baseline systems are the ones described in Section 2.3, and the last two are arguably more sophisticated clustering algorithms that are employed in an attempt to strengthen our baseline results.

### 4.2.1 Clustering Using the Second Eigenvector Only

As our first baseline, we adopt Shi and Malik's (2000) approach and cluster the reviews using only the second eigenvector, $\mathbf{e}_2$, as described in Section 2.3. Results on POL and the sentiment datasets, expressed in terms of accuracy and ARI, are shown in row 1 of Tables 2a and 2b, respectively. Owing to the randomness in the choice of seeds for 2-means, these and all other experimental results involving 2-means are averaged over ten independent runs.[11]

As we can see, this baseline achieves an accuracy of 93.7% on POL, but much lower accuracies (of 50–70%) on the sentiment datasets. The same performance trend can be observed with ARI. These results provide suggestive evidence that producing a sentiment-based clustering requires different features than producing a topic-based clustering, and that in many cases, the more salient features tend to be topic-based. The difference between sentiment-based clustering and topic-based clustering will be further illuminated by the experiments in Section 4.7.

In addition, it is worth noting that this baseline achieves much lower accuracies and ARI values on BOO, DVD, and ELE than on the remaining two sentiment datasets. Since $\mathbf{e}_2$ captures the most prominent dimension, these results suggest that sentiment dimension is not the most prominent dimension in these three datasets. In fact, this is intuitively plausible. For instance, in the book domain, positive book reviews typically contain a short description of the content, with the reviewer only briefly expressing her sentiment somewhere in the review. Similarly for the electronics domain: electronic product reviews are typically aspect-oriented, with the reviewer talking about the pros and cons of each aspect of the product (e.g., battery, durability). Since the reviews are likely to contain both positive and negative sentiment-bearing words, the sentiment-based clustering is unlikely to be captured by $\mathbf{e}_2$.

---

11. Note that clustering in a one-dimensional space (as in this baseline) yields very stable results regardless of the choice of seeds: our results over the ten runs exhibit nearly zero variance.

| | Accuracy | | | | | |
|---|---|---|---|---|---|---|
| System Variation | POL | MOV | KIT | BOO | DVD | ELE |
| 2nd eigenvector only | 93.7 | 70.9 | **69.7** | 58.9 | 55.3 | 50.8 |
| Top five eigenvectors | 95.9 | 58.9 | 64.0 | 59.9 | 60.4 | 63.8 |
| Interested Reader Model | **98.7** | 61.8 | 62.2 | 52.5 | 50.6 | 50.2 |
| NMF | 70.3 | **71.3** | 66.9 | 52.1 | 50.3 | 63.8 |
| Our system | 93.7 | 70.9 | **69.7** | **69.5** | **70.8** | **65.8** |

(a)

| | Adjusted Rand Index (ARI) | | | | | |
|---|---|---|---|---|---|---|
| System Variation | POL | MOV | KIT | BOO | DVD | ELE |
| 2nd eigenvector only | 0.76 | 0.17 | **0.15** | 0.03 | 0.01 | 0.01 |
| Top five eigenvectors | 0.84 | 0.03 | 0.05 | 0.04 | 0.03 | 0.07 |
| Interested Reader Model | **0.94** | 0.05 | 0.06 | 0.01 | 0.01 | 0.01 |
| NMF | 0.16 | **0.18** | 0.11 | 0.01 | 0.01 | 0.08 |
| Our system | 0.76 | 0.17 | **0.15** | **0.15** | **0.17** | **0.10** |

(b)

Table 2: Results in terms of (a) accuracy and (b) Adjusted Rand Index for the six datasets obtained using the bag-of-words document representation. The strongest result(s) for each dataset are boldfaced.

### 4.2.2 Clustering Using the Top Five Eigenvectors

As our second baseline, we represent each data point using the top five eigenvectors (i.e., $\mathbf{e}_1$ through $\mathbf{e}_5$), and cluster them using 2-means in this five-dimensional space, as described in Section 2.3. Hence, this can be thought of as an "ensemble" approach, where the clustering decision is collectively made by the five eigenvectors.[12]

Results are shown in row 2 of Tables 2a and 2b.[13] In comparison to the first baseline, we see improvements in accuracy and ARI for POL and the three sentiment datasets on which the first baseline performs poorly (i.e., BOO, DVD, and ELE), with the most drastic improvement observed on ELE. However, performance on the remaining two sentiment datasets deteriorates. These results can be attributed to the fact that for BOO, DVD, and ELE, $\mathbf{e}_2$ does not capture the sentiment dimension, but since some other eigenvector in the ensemble does, we see improvements. On the other hand, $\mathbf{e}_2$ has already captured the sentiment dimension in MOV and KIT; as a result, employing additional dimensions, which may not be sentiment-related, may only introduce noise into the computation of the similarities between the reviews.

---

12. While the first eigenvector can only produce a trivial clustering in which all data points reside in the same cluster, it is commonly used in combination with other top eigenvectors to create a low-dimensional space in which data points are clustered. See the work of Ng et al. (2001) for more details.

13. When clustering in a five-dimensional space, we observe that the results can be highly sensitive to the choice of seeds. For instance, the variances in the accuracy observed over the ten runs for POL, MOV, KIT, BOO, DVD, and ELE are 0, 2.38, 19.90, 24.70, 12.76, and 4.43, respectively.

### 4.2.3 CLUSTERING USING THE INTERESTED READER MODEL

Our third baseline is Kamvar et al.'s (2003) unsupervised clustering algorithm, which, according to the authors, is ideally suited for text clustering, and has recently been proved to be a special case of ratio-cut optimization (Kulis, Basu, Dhillon, & Mooney, 2009). Specifically, they introduce a new Laplacian inspired by the "Interested Reader Model". This Laplacian is computed as $(S + d_{max}I - D)/d_{max}$, where $D$ and $S$ are defined as in Section 2.1, except that $S_{i,j}{=}0$ if $i$ is not one of $j$'s $k$ nearest neighbors and $j$ is not one of $i$'s $k$ nearest neighbors; $d_{max}$ is the maximum rowsum of $S$; and $I$ is the identity matrix. Since its performance is highly sensitive to $k$, we tested values of 10, 15, ..., 500 for $k$ and report in row 3 of Tables 2a and 2b the *best* results. Somewhat disappointingly, despite its algorithmic sophistication and the fact that we are reporting the best results, this baseline does not offer consistent improvements over the previous two. In comparison to the first baseline, it achieves better performance on POL but worse performance on all the sentiment datasets. Like the first baseline, its results on BOO, DVD and ELE are particularly poor.

### 4.2.4 CLUSTERING USING NON-NEGATIVE MATRIX FACTORIZATION

Non-negative matrix factorization (NMF) has recently been shown by Xu, Liu, and Gong (2003) to be effective for document clustering. After re-implementing this algorithm, we evaluate it on our six datasets.[14] Shown in row 4 of Tables 2a and 2b are the *best* results obtained after running the algorithm five times. In comparison to the first baseline, NMF achieves better performance on ELE, comparable performance on MOV, and worse performance on the remaining datasets.

## 4.3 Our Active Clustering Algorithm

In this subsection, we describe human and automatic experiments for evaluating our active clustering algorithm.

### 4.3.1 HUMAN EXPERIMENTS

Unlike the four baselines, our active clustering algorithm requires users to specify which of the four dimensions (defined by the second through fifth eigenvectors) are most closely related to sentiment by inspecting a set of features derived from the unambiguous reviews for each dimension using MMFR. To better understand how easy it is for a human to select the desired dimension given the features, we performed the experiment independently with five humans (all of whom are computer science graduate students not affiliated with this research) and computed the agreement rate.

Specifically, for each dataset, we showed each human judge the top 100 features for each cluster according to MMFR (see Tables 3–8 for a subset of these 100 features induced for each of the six datasets, where the lightly shared columns correspond to the sentiment dimension selected by the majority of the human judges).[15] In addition, we informed her

---

14. For matrix factorization we use the code downloaded from http://www.csie.ntu.edu.tw/~cjlin/nmf/index.html.
15. While all human judges reported that inspecting the top 100 features is sufficient for identifying the sentiment dimension, we note that a user of our clustering algorithm may request to inspect as many features as she wants.

| POL | | | |
|---|---|---|---|
| $\mathbf{e}_2$ | $\mathbf{e}_3$ | $\mathbf{e}_4$ | $\mathbf{e}_5$ |
| $\mathbf{C}_1$ | $\mathbf{C}_1$ | $\mathbf{C}_1$ | $\mathbf{C}_1$ |
| serder | beyer | serbs | escrow |
| armenian | arabs | palestinians | serial |
| turkey | andi | muslims | algorithm |
| armenians | research | wrong | chips |
| muslims | israelis | department | ensure |
| sdpa | tim | bosnia | care |
| argic | uci | live | strong |
| davidian | ab | matter | police |
| dbd@ura | z@virginia | freedom | omissions |
| troops | holocaust | politics | excepted |
| | | | |
| $\mathbf{C}_2$ | $\mathbf{C}_2$ | $\mathbf{C}_2$ | $\mathbf{C}_2$ |
| sternlight | escrow | standard | internet |
| wouldn | sternlight | sternlight | uucp |
| pgp | algorithm | des | uk |
| crypto | access | escrow | net |
| algorithm | net | employer | quote |
| isn | des | net | ac |
| likely | privacy | york | co |
| access | uk | jake | didn |
| idea | systems | code | ai |
| cryptograph | pgp | algorithm | mit |

Table 3: Top ten features induced for each dimension for the POL domain. The shaded columns correspond to the dimensions selected by the human judges. $\mathbf{e}_2$, ..., $\mathbf{e}_5$ are the top eigenvectors; $\mathbf{C}_1$ and $\mathbf{C}_2$ are the clusters.

| MOV | | | |
|-----|-----|-----|-----|
| $\mathbf{e}_2$ | $\mathbf{e}_3$ | $\mathbf{e}_4$ | $\mathbf{e}_5$ |
| $\mathbf{C}_1$ | $\mathbf{C}_1$ | $\mathbf{C}_1$ | $\mathbf{C}_1$ |
| relationship | production | jokes | starts |
| son | earth | kids | person |
| tale | sequences | live | saw |
| husband | aliens | animation | feeling |
| perfect | war | disney | lives |
| drama | crew | animated | told |
| focus | alien | laughs | happen |
| strong | planet | production | am |
| beautiful | horror | voice | felt |
| nature | evil | hilarious | happened |
| $\mathbf{C}_2$ | $\mathbf{C}_2$ | $\mathbf{C}_2$ | $\mathbf{C}_2$ |
| worst | sex | thriller | comic |
| stupid | romantic | killer | sequences |
| waste | school | murder | michael |
| bunch | relationship | crime | supporting |
| wasn | friends | police | career |
| video | jokes | car | production |
| worse | laughs | dead | peter |
| boring | sexual | killed | style |
| guess | cute | starts | latest |
| anyway | mother | violence | entertaining |

Table 4: Top ten features induced for each dimension for the MOV domain. The shaded columns correspond to the dimensions selected by the human judges. $\mathbf{e}_2, \ldots, \mathbf{e}_5$ are the top eigenvectors; $\mathbf{C}_1$ and $\mathbf{C}_2$ are the clusters.

| BOO | | | |
|:---:|:---:|:---:|:---:|
| $\mathbf{e}_2$ | $\mathbf{e}_3$ | $\mathbf{e}_4$ | $\mathbf{e}_5$ |
| $\mathbf{C}_1$ | $\mathbf{C}_1$ | $\mathbf{C}_1$ | $\mathbf{C}_1$ |
| history | series | loved | must |
| must | man | highly | wonderful |
| modern | history | easy | old |
| important | character | enjoyed | feel |
| text | death | children | away |
| reference | between | again | children |
| excellent | war | although | year |
| provides | seems | excellent | someone |
| business | political | understand | man |
| both | american | three | made |
| | | | |
| $\mathbf{C}_2$ | $\mathbf{C}_2$ | $\mathbf{C}_2$ | $\mathbf{C}_2$ |
| plot | buy | money | boring |
| didn | bought | bad | series |
| thought | information | nothing | history |
| boring | easy | waste | pages |
| got | money | buy | information |
| character | recipes | anything | between |
| couldn | pictures | doesn | highly |
| ll | look | already | page |
| ending | waste | instead | excellent |
| fan | copy | seems | couldn |

Table 5: Top ten features induced for each dimension for the BOO domain. The shaded columns correspond to the dimensions selected by the human judges. $\mathbf{e}_2$, ..., $\mathbf{e}_5$ are the top eigenvectors; $\mathbf{C}_1$ and $\mathbf{C}_2$ are the clusters.

| ELE | | | |
|---|---|---|---|
| $\mathbf{e}_2$ | $\mathbf{e}_3$ | $\mathbf{e}_4$ | $\mathbf{e}_5$ |
| $\mathbf{C}_1$ | $\mathbf{C}_1$ | $\mathbf{C}_1$ | $\mathbf{C}_1$ |
| mouse | music | easy | amazon |
| cable | really | used | cable |
| cables | ipod | card | card |
| case | too | fine | recommend |
| red | little | using | dvd |
| monster | headphones | problems | camera |
| picture | hard | fine | fast |
| kit | excellent | drive | far |
| overall | need | computer | printer |
| paid | fit | install | picture |
| | | | |
| $\mathbf{C}_2$ | $\mathbf{C}_2$ | $\mathbf{C}_2$ | $\mathbf{C}_2$ |
| working | worked | money | phone |
| never | problem | worth | off |
| before | never | amazon | worked |
| phone | item | over | power |
| days | amazon | return | battery |
| headset | working | years | unit |
| money | support | much | set |
| months | months | headphones | phones |
| return | returned | sony | range |
| second | another | received | little |

Table 6: Top ten features induced for each dimension for the ELE domain. The shaded columns correspond to the dimensions selected by the human judges. $\mathbf{e}_2$, …, $\mathbf{e}_5$ are the top eigenvectors; $\mathbf{C}_1$ and $\mathbf{C}_2$ are the clusters.

| KIT | | | |
|---|---|---|---|
| **$e_2$** | **$e_3$** | **$e_4$** | **$e_5$** |
| **$C_1$** | **$C_1$** | **$C_1$** | **$C_1$** |
| love | works | really | pan |
| clean | water | nice | oven |
| nice | clean | works | cooking |
| size | work | too | made |
| set | ice | quality | pans |
| kitchen | makes | small | better |
| easily | thing | sturdy | heat |
| sturdy | need | little | cook |
| recommend | keep | think | using |
| price | best | item | clean |
| | | | |
| **$C_2$** | **$C_2$** | **$C_2$** | **$C_2$** |
| months | price | ve | love |
| still | item | years | coffee |
| back | set | love | too |
| never | ordered | never | recommend |
| worked | amazon | clean | makes |
| money | gift | months | over |
| did | got | over | size |
| amazon | quality | pan | little |
| return | received | been | maker |
| machine | knives | pans | cup |

Table 7: Top ten features induced for each dimension for the KIT domain. The shaded columns correspond to the dimensions selected by the human judges. $e_2$, ..., $e_5$ are the top eigenvectors; $C_1$ and $C_2$ are the clusters.

| DVD | | | |
|:---:|:---:|:---:|:---:|
| $\mathbf{e}_2$ | $\mathbf{e}_3$ | $\mathbf{e}_4$ | $\mathbf{e}_5$ |
| $\mathbf{C}_1$ | $\mathbf{C}_1$ | $\mathbf{C}_1$ | $\mathbf{C}_1$ |
| worth | music | video | money |
| bought | collection | music | quality |
| series | excellent | found | video |
| money | wonderful | feel | worth |
| season | must | bought | found |
| fan | loved | workout | version |
| collection | perfect | daughter | picture |
| music | highly | recommend | waste |
| tv | makes | our | special |
| thought | special | disappointed | sound |
| | | | |
| $\mathbf{C}_2$ | $\mathbf{C}_2$ | $\mathbf{C}_2$ | $\mathbf{C}_2$ |
| young | worst | series | saw |
| between | money | cast | watched |
| actors | thought | fan | loved |
| men | boring | stars | enjoy |
| cast | nothing | original | whole |
| seems | minutes | comedy | got |
| job | waste | actors | family |
| beautiful | saw | worth | series |
| around | pretty | classic | season |
| director | reviews | action | liked |

Table 8: Top ten features induced for each dimension for the DVD domain. The shaded columns correspond to the dimensions selected by the human judges. $\mathbf{e}_2, \ldots, \mathbf{e}_5$ are the top eigenvectors; $\mathbf{C}_1$ and $\mathbf{C}_2$ are the clusters.

| Judge | POL | MOV | KIT | BOO | DVD | ELE |
|-------|-----|-----|-----|-----|-----|-----|
| 1 | 2,3,4 | 2 | 2 | 4 | 3 | 3 |
| 2 | 2,4 | 2 | 2 | 4 | 3 | 3 |
| 3 | 4 | 2,4 | 4 | 4 | 3 | 3 |
| 4 | 2,3 | 2 | 2 | 4 | 3 | 3,4 |
| 5 | 2 | 2 | 2 | 4 | 3 | 3 |
| Agreement | 80% | 100% | 80% | 100% | 100% | 100% |

Table 9: Human agreement rate. Also shown are the eigenvectors selected by the five judges.

of the intended dimension: for example, for POL, the judge was told that the intended clustering was Politics vs. Science. Also, if she determined that more than one dimension was relevant to the intended clustering, she was instructed to rank these dimensions in terms of relevance, where the most relevant one would appear first in the list.

The dimensions (expressed in terms of the IDs of the eigenvectors) selected by each of the five judges for each dataset are shown in Table 9. The agreement rate (shown in the last row of the table) was computed based on only the highest-ranked dimension selected by each judge. As we can see, perfect agreement is achieved for four of the five sentiment datasets, and for the remaining two datasets, near-perfect agreement is achieved. These results, together with the fact that it took five to six minutes to identify the relevant dimension, indicate that asking a human to determine the intended dimension based on solely the "informative" features is a viable task.

### 4.3.2 CLUSTERING RESULTS

Next, we cluster all 2000 documents for each dataset using the dimension selected by the majority of the human judges. The clustering results are shown in row 5 of Tables 2a and 2b. In comparison to the best baseline for each dataset, we see that our algorithm performs substantially better on BOO, DVD and ELE, at almost the same level on MOV and KIT, but slightly worse on POL. Note that the improvements observed for BOO, DVD and ELE can be attributed to the failure of $e_2$ to capture the sentiment dimension. Perhaps most importantly, by exploiting human feedback, our algorithm has achieved more stable performance across the datasets than the four baselines.[16]

### 4.3.3 IDENTIFICATION OF UNAMBIGUOUS DOCUMENTS

Recall that the features with the largest MMFR were computed from the unambiguous documents only. To get an idea of how accurate our algorithm for identifying unambiguous documents is, we show in Table 10 the accuracy obtained when the unambiguous documents in each dataset were clustered using the eigenvector selected by the majority of the judges. As we can see, the accuracy on each dataset is higher than the corresponding accuracy shown in row 5 of Table 2a. In fact, an accuracy of more than 85% was achieved on all but one dataset. This suggests that our method of identifying unambiguous documents is reasonably accurate.

---

16. As in the first baseline, since we are clustering in a one-dimensional space here, the results are not sensitive to the choice of seeds, yielding zero variance over the ten independent runs.

|  | POL | MOV | KIT | BOO | DVD | ELE |
|---|---|---|---|---|---|---|
| Accuracy | 99.8 | 87.0 | 87.6 | 86.2 | 87.4 | 77.6 |

Table 10: Accuracies on unambiguous documents.

|  | POL | MOV | KIT | BOO | DVD | ELE |
|---|---|---|---|---|---|---|
| # labels | 400 | 150 | 200 | 350 | 350 | 200 |

Table 11: Transductive SVM results.

Note that it is crucial to be able to achieve a high accuracy on the unambiguous documents: if clustering accuracy is low, the features induced from the clusters may not be an accurate representation of the corresponding dimension, and the human judge may have a difficult time identifying the intended dimension. In fact, some human judges reported difficulty in identifying the correct dimension for the ELE dataset, and this can be attributed in part to the low accuracy achieved on the unambiguous documents.

### 4.3.4 USER FEEDBACK VERSUS LABELED DATA

Recall that our four baselines are unsupervised, whereas our algorithm can be characterized as semi-supervised, as it relies on user feedback to select the intended dimension. Hence, it should not be surprising to see that the average clustering performance of our algorithm is better than that of the baselines.

To do a fairer comparison, we conduct another experiment in which we compare our algorithm against a semi-supervised sentiment classification system, which uses a transductive SVM as the underlying semi-supervised learner. More specifically, the goal of this experiment is to determine how many labeled documents are needed in order for the transductive learner to achieve the same level of performance as our algorithm. To answer this question, we first give the transductive learner access to the 2000 documents for each dataset as unlabeled data. Next, we randomly sample 50 unlabeled documents and assign them the true label. We then re-train the classifier and compute its accuracy on the 2000 documents. We keep adding more labeled data (50 in each iteration) until it reaches the accuracy achieved by our algorithm. Results of this experiment are shown in Table 11. Owing in the randomness involved in the selection of unlabeled documents, these results are averaged over ten independent runs. As we can see, our user feedback is equivalent to the effort of hand-annotating 275 documents per dataset on average.

### 4.3.5 MULTIPLE RELEVANT EIGENVECTORS

As seen from Table 9, some human judges selected more than one eigenvector for some datasets (e.g., {2,3,4} for POL; {2,4} for MOV; and {3,4} for ELE). However, we never took into account these "extra" eigenvectors in our previous experiments. To better understand whether these extra eigenvectors can help improve accuracy and ARI, we conduct another experiment in which we apply 2-means to cluster the documents in the space defined by all of the selected eigenvectors. Table 12 shows the accuracy and ARI results that are averaged over ten independent runs. As we can see, the results for POL are considerably better than those obtained when only the highest-ranked eigenvector is used, suggesting that the extra eigenvectors contain useful information. However, the results on MOV and ELE drop slightly with the addition of the extra eigenvectors, indicating that the extra sentiment dimensions are not useful.

| | POL | | MOV | | ELE | |
|---|---|---|---|---|---|---|
| | Acc | ARI | Acc | ARI | Acc | ARI |
| Our system | 95.9 | 0.84 | 69.1 | 0.16 | 65.1 | 0.10 |

Table 12: Results obtained using multiple relevant eigenvectors for the POL, MOV and ELE datasets.

| | POL | MOV | KIT | BOO | DVD | ELE |
|---|---|---|---|---|---|---|
| Accuracy | 99.3 | 86.1 | 81.7 | 79.3 | 77.6 | 80.6 |

Table 13: Supervised classification accuracies.

### 4.3.6 SUPERVISED CLASSIFICATION RESULTS

Next, we present results for supervised classification on our five sentiment datasets. While one should not expect our largely unsupervised approach to offer comparable performance to a fully-supervised approach, we believe that having fully-supervised results will enable the reader to get a sense of where our work stands among existing work on identifying sentiment in these datasets. Specifically, we report in Table 13 averaged 10-fold cross-validation accuracies, where an SVM classifier is trained on nine folds and tested on the remaining fold in each fold experiment. As we can see, our results lag behind the supervised results by 8.1–15.2% on these datasets.

## 4.4 Alternative Document Representations

In the above experiments, we represented each document as a bag of words with the most frequent 1.5% of the words removed. This is, of course, not the only way to represent a document. In this subsection, we examine two alternative document representations in an attempt to better understand the effect of document representation on classification results.

In our first document representation, we represent a document using all of the unigrams that appear in it and do *not* remove the frequent words from the document vector. This *bag-of-all-words* (BOAW) representation is motivated by the fact that the frequencies of function words and the like have been shown in many studies to be useful features for various kinds of non-topic-based classification (e.g., Finn & Kushmerick, 2006; Stein, Argamon, & Frieder, 2006; Abbasi, Chen, & Salem, 2008; Koppel, Schler, & Argamon, 2009). The

| | Accuracy | | | | | |
|---|---|---|---|---|---|---|
| System Variation | POL | MOV | KIT | BOO | DVD | ELE |
| 2nd eigenvector only | 70.6 | 54.3 | 51.6 | 52.4 | 51.2 | 53.1 |
| Top five eigenvectors | **94.7** | 60.6 | 58.0 | 56.1 | 53.7 | 57.1 |
| Interested Reader Model | 61.2 | 61.1 | 57.8 | 52.4 | 50.4 | 50.3 |
| NMF | 59.2 | 54.6 | 50.8 | 50.1 | 52.9 | 51.4 |
| Our system | 84.3 | **65.9** | **64.8** | **60.1** | **58.6** | **64.1** |

(a)

| | Adjusted Rand Index (ARI) | | | | | |
|---|---|---|---|---|---|---|
| System Variation | POL | MOV | KIT | BOO | DVD | ELE |
| 2nd eigenvector only | 0.17 | 0.01 | 0.01 | 0.01 | 0.01 | 0.01 |
| Top five eigenvectors | **0.80** | 0.04 | 0.03 | 0.01 | 0.01 | 0.03 |
| Interested Reader Model | 0.05 | 0.05 | 0.02 | 0.01 | 0.01 | 0.01 |
| NMF | 0.03 | 0.01 | −0.01 | −0.01 | 0.01 | 0.01 |
| Our system | 0.47 | **0.10** | **0.09** | **0.04** | **0.03** | **0.08** |

(b)

Table 14: Results in terms of (a) accuracy and (b) Adjusted Rand Index for the six datasets obtained using the bag-of-all-words document representation. The strongest result(s) for each dataset are boldfaced.

accuracy and ARI results obtained by re-running our four baselines as well as our system using this document representation are shown in Tables 14a and 14b, respectively. Comparing Tables 2a and 14a, we can see that when all words are used as features, the best accuracy achieved for each dataset drops by 3–11% than when the high-frequency words are removed before spectral clustering is applied. Similar trends can be observed with the ARI results shown in Tables 2b and 14b. Overall, these results substantiate our hypothesis that retaining the high-frequency words in the document representation has an adverse effect on the performance of these clustering algorithms.

Next, we experiment with another representation, specifically one in which each document is represented using only the sentiment-bearing words it contains. To understand the motivation behind this *bag-of-sentiment-words* (BOSW) representation, recall from the introduction that one way to "encourage" the clustering algorithm to produce the user-desired clustering is to design the feature space so that it contains all and only those features that are useful for producing the user-desired clustering. Since we desire a sentiment-based clustering, we can design a feature space composed of solely sentiment-bearing words. Since a hand-crafted *subjectivity lexicon* (i.e., a lexicon where each word is manually labeled with its prior polarity[17]) for English is readily available, we can automatically construct a feature space that consists of only those words that have a (positive or negative) polarity according to the subjectivity lexicon, and represent a document using the resulting feature space. The

---

17. The prior polarity of a word is its polarity computed without regard to the context in which the word appears.

| | Accuracy | | | | |
|---|---|---|---|---|---|
| System Variation | MOV | KIT | BOO | DVD | ELE |
| 2nd eigenvector only | **69.1** | **62.3** | **60.2** | 61.4 | **63.9** |
| Top five eigenvectors | 60.7 | 57.9 | 57.6 | 63.1 | 62.7 |
| Interested Reader Model | 54.6 | 50.3 | 54.4 | 56.0 | 50.6 |
| NMF | 68.8 | 59.0 | 59.2 | **63.3** | 60.5 |
| Our system | **69.1** | **62.3** | **60.2** | 61.4 | **63.9** |

(a)

| | Adjusted Rand Index (ARI) | | | | |
|---|---|---|---|---|---|
| System Variation | MOV | KIT | BOO | DVD | ELE |
| 2nd eigenvector only | **0.15** | **0.06** | **0.04** | 0.05 | **0.08** |
| Top five eigenvectors | 0.04 | 0.03 | 0.03 | **0.07** | 0.06 |
| Interested Reader Model | 0.01 | 0.01 | 0.01 | 0.01 | 0.01 |
| NMF | 0.14 | 0.03 | 0.03 | **0.07** | 0.04 |
| Our system | **0.15** | **0.06** | **0.04** | 0.05 | **0.08** |

(b)

Table 15: Results in terms of (a) accuracy and (b) Adjusted Rand Index for the five sentiment datasets obtained using the bag-of-sentiment-words document representation. The strongest result(s) for each dataset are boldfaced.

goal, then, is to determine whether the BOSW document representation can improve the sentiment-based clustering results obtained using the BOW representation.

To identify sentiment-bearing words in our experiment, we employ the subjectivity lexicon introduced in the work of Wilson, Wiebe, and Hoffmann (2005).[18] The lexicon contains 8221 words, each of which is hand-labeled with a prior polarity of POSITIVE, NEGATIVE, or NEUTRAL. We create a new subjectivity lexicon $L$ in which we retain only those words in Wilson et al.'s lexicon that have either a POSITIVE or NEGATIVE polarity. The BOSW representation of a document is then composed of all and only those words that appear in both $L$ and the document.

The accuracy and ARI results of our baselines and our system obtained when employing the BOSW representation are shown in Tables 15a and 15b, respectively. Consider first the "second eigenvector only" baseline, NMF, and the Interested Reader Model. In comparison to their corresponding results in Tables 2a and 2b, where the BOW representation was used, we can see that performance improves on the BOO, DVD, and ELE datasets in most cases, but drops on the MOV and KIT datasets. For the "top five eigenvectors" baseline, performance increases on DVD and slightly on MOV, but drops on the remaining datasets. Finally, using the BOSW representation causes the performance of our system to drop on all datasets.

Overall, these results seem to suggest that whether the BOSW representation of a document yields better clustering results than its BOW representation is rather dependent on the underlying domain and clustering algorithm. Nevertheless, we can see that the best clustering accuracy/ARI achieved for each sentiment dataset using the BOSW representation is significantly lower than that obtained using the BOW representation. We speculate two reasons for the poorer results. First, the general-purpose subjectivity lexicon does not cover all of the sentiment-bearing words. In particular, words that are sentiment-oriented in the context of a particular domain but have a neutral polarity otherwise may be omitted from the BOSW document representation. Second, some non-sentiment-bearing words might be useful for identifying sentiment.

---

18. See http://www.cs.pitt.edu/mpqa/.

## 4.5 Domain Adaptation

As mentioned in the introduction, the majority of existing approaches to sentiment classification is supervised. One weakness of these supervised approaches is that when given a new domain, one needs to go through the expensive process of collecting a large amount of annotated data in order to train an accurate polarity classifier.[19] One may argue that our active clustering algorithm suffers from the same weakness: the user needs to identify the sentiment dimension for each domain. One way to address this weakness is through domain adaptation. Specifically, we investigate whether the sentiment dimension manually identified for one domain (henceforth the *source* domain) can be used to *automatically* identify the sentiment dimension for a new domain (henceforth the *target* domain). We hypothesize that domain adaptation is feasible, especially if the two domains are *sentimentally similar* (i.e., there is a significant overlap between the features that characterize the sentiment dimensions of the two domains).

As a result, we propose the following method for automatically identifying the sentiment dimension for the target domain, $y$, using the sentiment dimension manually identified for the source domain, $x$. Assume that the sentiment dimension of domain $x$ is defined by eigenvector $\mathbf{e}^x$. Moreover, assume that $C_1^{\mathbf{e}^x}$ and $C_2^{\mathbf{e}^x}$ are the two vectors of the top-ranked features (obtained using MMFR) that characterize the two clusters induced by $\mathbf{e}^x$ (with 100 features in each cluster). Now, given the target domain $y$, we first compute the similarity between $\mathbf{e}^x$ and each of $y$'s top eigenvectors, $\mathbf{e}_2^y$, ..., $\mathbf{e}_5^y$, where the similarity between two eigenvectors $\mathbf{e}^x$ and $\mathbf{e}^y$ is defined as

$$\max(\phi(C_1^{\mathbf{e}^x}, C_1^{\mathbf{e}^y}) + \phi(C_2^{\mathbf{e}^x}, C_2^{\mathbf{e}^y}), \phi(C_1^{\mathbf{e}^x}, C_2^{\mathbf{e}^y}) + \phi(C_2^{\mathbf{e}^x}, C_1^{\mathbf{e}^y}))$$

Here, $\phi$ is a similarity function that computes the similarity between two feature vectors. In our experiments, we simply set it to be the dot product, which allows us to capture the degree of overlap between the two feature vectors. Then, we posit the eigenvector from $\{\mathbf{e}_2^y, \ldots, \mathbf{e}_5^y\}$ that has the highest overlap as the one that defines the sentiment dimension.[20]

To determine the effectiveness of our method, we compare the automatically selected eigenvector with the human-selected eigenvector for each domain. Results are shown in Table 16, where a 'Y' in row $i$ and column $j$ indicates that the sentiment dimension for target domain $j$ has been successfully identified by using the sentiment dimension manually identified for source domain $i$, and an 'N' indicates a failure. For instance, if we know the sentiment dimension of the DVD domain (through human feedback), then our domain adaptation method can be used to correctly identify the sentiment domain of MOV and vice versa. However, domain adaptation using our method is not always successful. For instance, knowing the sentiment dimension of MOV does not allow us to correctly predict the sentiment dimension of ELE. Interestingly, if we ignore the BOO/KIT pair, then domain adaptation exhibits symmetry. By symmetry, we mean that if domain $x$ can be used to identify the correct sentiment dimension for domain $y$, then domain $y$ can be used to identify the correct sentiment dimension for domain $x$. This intuitively makes sense: if $x$ can successfully be used to identify the sentiment dimension for $y$, it is likely that the two domains share a lot of sentiment words. Consequently, using $y$ to adapt to $x$ is also likely to be successful. The BOO/KIT pair represents a case in which domain adaptation is successful in only one direction: while domain adaptation is successful from BOO to KIT, the similarity between the sentiment dimensions of the two domains is not high (see the discussion in the next paragraph for details), which contributes to the failure to adaptation in the other direction.

---

19. While collecting annotated data is trivial when dealing with review data, the same is not necessarily true for other kinds of data. For instance, people express their opinions and sentiment in political blogs and floor debates, but the associated postings and transcripts may not be explicitly annotated with sentiment labels.

20. Note that the two arguments to the max function correspond to the two different ways of creating the mapping between the feature vectors in the two domains.

| Domain | MOV | DVD | BOO | ELE | KIT |
|--------|-----|-----|-----|-----|-----|
| MOV | – | Y | N | N | N |
| DVD | Y | – | Y | N | Y |
| BOO | N | Y | – | N | Y |
| ELE | N | N | N | – | Y |
| KIT | N | Y | N | Y | – |

Table 16: Domain adaptation results.

As mentioned at the beginning of this subsection, we hypothesize that domain adaptation is likely to be successful if the two domains under consideration are "similar" to each other. To test this hypothesis, we show in Table 17a the similarity between the manually identified eigenvector and the corresponding automatically identified eigenvector for each pair of domains. Three points deserve mention. First, as long as the similarity value is at least 14, domain adaptation is successful; also, as long as the similarity value is at most 6, domain adaptation is unsuccessful. Hence, these results substantiate our hypothesis that domain adaptation is more likely to be successful if the two domains under consideration are more similar to each other. It would be interesting to see if these two thresholds can be used to predict whether domain adaptation is successful given a new pair of domains. Second, domain adaptation in both directions are likely to be successful if the similarity value is "sufficiently" high. As mentioned before, if the similarity value is high, then the two domains share many sentiment words in common, which may in turn contribute to successful domain adaptation in both directions. For the five domains we are considering, as long as the similarity value is at least 14, then domain adaptation in both directions will be successful. Third, it is worth reiterating that even if the similarity value falls below this threshold, it does not imply that domain adaptation will fail. As mentioned before, the sentiment dimension for domain $y$ will be (correctly) identified as long as its similarity with

| Domain | MOV | DVD | BOO | ELE | KIT |
|--------|-----|-----|-----|-----|-----|
| MOV | – | 14 | (6) | (2) | (3) |
| DVD | 14 | – | 21 | (10) | 10 |
| BOO | (6) | 21 | – | (10) | 8 |
| ELE | (3) | (8) | (6) | – | 32 |
| KIT | (1) | 10 | (11) | 32 | – |

(a)

| Domain | MOV | DVD | BOO | ELE | KIT |
|--------|-----|-----|-----|-----|-----|
| MOV | – | 10 | (4) | (2) | (3) |
| DVD | 5 | – | 13 | (9) | 7 |
| BOO | (4) | 14 | – | (6) | 6 |
| ELE | (2) | (8) | (5) | – | 23 |
| KIT | (1) | 7 | (8) | 27 | – |

(b)

| Domain | MOV | DVD | BOO | ELE | KIT |
|--------|-----|-----|-----|-----|-----|
| MOV | – | 4 | (2) | (0) | (0) |
| DVD | 9 | – | 8 | (1) | 3 |
| BOO | (2) | 7 | – | (4) | 2 |
| ELE | (0) | (0) | (1) | – | 9 |
| KIT | (1) | 3 | (3) | 5 | – |

(c)

Table 17: Similarity results for domain adaptation. (a) shows the similarity between the sentiment eigenvector in the source domain and the eigenvector most similar to it in the target domain. (b) shows the similarity between the sentiment eigenvector in the source domain and the second most similar eigenvector in the target domain. (c) shows the similarity gap, which is the difference between the corresponding entries in (a) and (b).

the sentiment dimension for domain $x$ is highest among the four eigenvectors for $y$, as is the case with the BOO/KIT domain pair.

So far we have attempted to correlate the success of domain adaptation with the similarity between the manually selected eigenvector in the source domain and the eigenvector most similar to it in the target domain. It may be worth to also consider the similarity between the manually selected eigenvector and the second most similar eigenvector in the target domain, as the *gap* in similarity may give an indication as to the success of domain adaptation. To determine whether there is a better correlation between the success of domain adaptation and this similarity gap, we compute (1) the similarity between the eigenvector manually selected for the source domain and its second most similar eigenvector in the target domain (see Table 17b) as well as (2) the similarity gap (see Table 17c), which is simply the difference between the corresponding entries in Tables 17a and 17b. As we can see from Table 17c, there also appears to be some correlation between the success of domain adaptation and the gap values. In particular, if the gap value is at least 5, domain adaptation is successful; however, if the gap value is at most 1, domain adaptation is unsuccessful. Nevertheless, these gap values do not help to predict the domain pairs where the success of domain adaptation cannot be predicted using the similarity values in Table 17a (e.g., the domain pairs that have low similarity and yet are domain-adaptable). Moreover, they fail to predict the success of domain adaptation for many domain pairs, specifically those where the gap value is between 1 and 5.

## 4.6 Subjectivity Lexicon versus Human Feedback

One might argue that if we had access to a subjectivity lexicon, we could use it to automatically identify the right sentiment dimension, thus obviating the need for human feedback altogether. In this subsection, we investigate whether it is indeed feasible to use a hand-built general-purpose sentiment lexicon to identify the eigenvector that corresponds to the sentiment dimension in a new domain.

For our experiment, we use the subjectivity lexicon $L$ described in Section 4.4. As mentioned before, $L$ contains all and only those words in Wilson et al.'s (2005) subjectivity lexicon that are marked with a prior polarity of POSITIVE or NEGATIVE. The procedure for automatically identifying the sentiment dimension using $L$ is similar to the one described in the domain adaptation section: for each of the second through fifth eigenvectors, we first compute the similarity between the eigenvector and $L$, and then choose the eigenvector that has the highest similarity with $L$. As in domain adaptation, we compute the similarity between $L$ and an eigenvector $e^x$ as

$$\max(\phi(C_1^L, C_1^{\mathbf{e}^x}) + \phi(C_2^L, C_2^{\mathbf{e}^x}), \phi(C_1^L, C_2^{\mathbf{e}^x}) + \phi(C_2^L, C_1^{\mathbf{e}^x}))$$

where $C_1^L$ and $C_2^L$ represent the words in $L$ that are labeled as positive and negative respectively, and $C_1^{\mathbf{e}^x}$ and $C_2^{\mathbf{e}^x}$ are the top-ranked features (obtained using MMFR) that characterize the two clusters induced by $\mathbf{e}^x$ (with 100 features in each cluster). $\phi$ is a similarity function that computes the similarity between two feature vectors. As in domain adaptation, we simply set it to be the dot product.

Our results indicate that we successfully identified the right eigenvector using $L$ for each of the five domains. Note that while $L$ is a general-purpose (i.e., domain-independent) lexicon containing only generic sentiment-bearing words, it is good enough to identify the correct sentiment dimension for five different domains. It is worth noting that the sentiment dimension of the MOV domain has the highest similarity with $L$ (i.e., 34) out of the five domains, suggesting that the highest-ranked sentiment features of the MOV domain (according to MMFR) are largely generic. DVD has the second largest similarity with $L$ (33), followed by BOO (26), KIT (16) and ELE (16). The comparatively low similarity values for KIT and ELE are indicative of the fact that their highest-ranked sentiment features are largely domain-specific.

Finally, although a subjectivity lexicon obviates the need for human feedback, we should emphasize that this does not undermine the contribution of our feedback-oriented clustering technique, for the following reasons. First, thinking from a text mining perspective, it would be good to have an approach that is as knowledge-free as possible. Employing a hand-crafted subjectivity lexicon makes our system resource-dependent; in fact, a subjectivity lexicon may not be readily available for the vast majority of natural languages. Second, we want our method to be potentially applicable to non-sentiment domains (e.g., spam vs. not spam), where we are again faced with the same problem that a hand-built lexicon may not be available.

## 4.7 Single Data, Multiple Clusterings

As mentioned previously, a set of documents can be clustered along different dimensions. For example, movie reviews can be clustered by *sentiment* (positive vs. negative) or *genre* (e.g., action, romantic or documentary). A natural question is: can we produce different clusterings of a given set of documents, each of which corresponds to a different dimension? For the vast majority of existing text clustering algorithms, the answer is no: they can only cluster along exactly one dimension, which is typically the most prominent dimension. On the other hand, since our algorithm induces the important clustering dimensions of a dataset, each of which can in principle be used to produce a (distinct) clustering, we hypothesize that it can generate multiple clusterings of a given dataset along its important dimensions.

To test our claim that our algorithm can produce multiple clusterings, we evaluate it on four datasets that possess multiple clustering dimensions, namely MOV-DVD, BOO-DVD, DVD-ELE, and MOV-KIT.[21] For example, the BOO-DVD dataset consists of all the reviews taken from the BOO and DVD domains. Hence, each "augmented" dataset is composed of 4000 reviews (2000 from each of the two contributing domains), which can be clustered according to either *topic* (e.g., Book vs. DVD) or *sentiment*.[22] Note that the four pairs of domains used to create the augmented datasets were chosen carefully. Specifically, two augmented datasets (MOV-DVD and BOO-DVD) were created such that their constituent domains are mutually domain-adaptable according to Table 16, and the remaining two (DVD-ELE and MOV-KIT) were created such that their constituent domains are not domain-adaptable. Our goal is to see whether our active clustering algorithm is able to produce both topic- and sentiment-based clusterings for datasets with different levels of sentimental similarity.

The clustering procedure is almost identical to the one described in Section 3. In essence, we (1) compute the top five eigenvectors of the Laplacian matrix; (2) learn the top-ranked features corresponding to $e_2$ through $e_5$ according to MMFR; (3) ask the human judges to identify the eigenvectors corresponding to both the topic dimension and the sentiment dimension; and (4) use 2-means to produce two clusterings of the reviews, one according to the selected topic dimension and the other the selected sentiment dimension. As in Section 4.3, we conducted human and automatic experiments to determine the viability of our algorithm.

---

21. The reason for our employing these augmented datasets is that they not only obviate the need for additional human annotations, but also guarantee that there are at least two dimensions along which clusters can be formed, thus allowing us to directly test its ability to produce multiple clusterings. While it is also possible to evaluate our algorithm's ability to generate multiple clusterings using the MOV dataset (by clustering along genre and sentiment), we decided to leave this for future investigation, since the documents in MOV are not annotated with genre information.

22. This is *not* to be confused with topic-sentiment mixture models (Mei, Ling, Wondra, Su, & Zhai, 2007), where the goal is to first use topic models to mine the major aspects of a product from an online review and then assign ratings to each extracted aspect. On the other hand, our goal is to design a clustering algorithm that is capable of generating multiple clusterings of a dataset.

| Judge | MOV-DVD | BOO-DVD | DVD-ELE | MOV-KIT |
|-------|---------|---------|---------|---------|
| 1 | 2 | 2 | 2 | 2 |
| 2 | 2 | 2 | 2 | 2 |
| 3 | 2 | 2 | 2 | 2 |
| 4 | 2 | 2 | 2 | 2 |
| 5 | 2 | 2 | 2 | 2 |
| Agreement | 100% | 100% | 100% | 100% |

(a)

| Judge | MOV-DVD | BOO-DVD | DVD-ELE | MOV-KIT |
|-------|---------|---------|---------|---------|
| 1 | 3 | 4,5 | 3 | 3 |
| 2 | 3,4 | 4,5 | 3,5 | 3,5 |
| 3 | 3,4 | 4,5 | 5,3 | 3 |
| 4 | 3 | 4,5 | 3 | 5 |
| 5 | 3 | 4,5 | 3 | 3 |
| Agreement | 100% | 100% | 80% | 80% |

(b)

Table 18: Human agreement rate for selecting (a) the topic dimension and (b) the sentiment dimension for the augmented datasets. Also shown are the eigenvectors selected by the human judges.

### 4.7.1 HUMAN EXPERIMENTS

We employed the same five human judges involved in the human experiments in Section 4.3 to independently determine the topic dimension and the sentiment dimension for each of the four augmented datasets using only the top features according to MMFR. As before, if a human judge identifies more than one relevant eigenvector for a particular dimension, we ask her to rank the eigenvectors according to relevance. Finally, we take the topic/sentiment dimension that is the ranked first by the largest number of judges as the human-selected topic/sentiment dimension.

Tables 18a and 18b show respectively the topic and sentiment dimensions (expressed in terms of the IDs of the eigenvectors) selected by each of the five judges for each augmented dataset. Also shown in the tables is the the human agreement rate, which was computed based on only the highest-ranked dimension selected by each judge. Several points from these human experiments deserve mention.

First, for each dataset, all human judges managed to find one eigenvector (out of the top five) that corresponds to topic *and* at least one other eigenvector that corresponds to sentiment. Perhaps more importantly, a human agreement rate of at least 80% was achieved on all four datasets with respect to selecting the eigenvector(s) that correspond to the topic and sentiment dimensions. These results together provide suggestive evidence that (1) the eigen-decomposition procedure in our active clustering algorithm is effective enough to unearth both the topic and sentiment dimensions when both of them are present in a dataset, and (2) our proposal for incorporating user feedback via inspecting a small number of features is viable.

Second, while both topic and sentiment are prominent dimensions in these datasets, the fact that the second eigenvector captures the topic dimension for all four datasets suggests that topic is a more prominent dimension than sentiment. In fact, all of our human judges reported that the topic dimension can be identified quite easily, achieving perfect agreement on identifying the topic dimension. This provides empirical evidence for our speculation that topic is typically (though not always) a more prominent dimension than sentiment when both dimensions exist in a dataset.

Third, while reasonably high human agreement rate for identifying the sentiment dimension was achieved (perfect agreement on two datasets and 80% agreement rate on the remaining two; see Table 18b for details), the human judges have reported that it was difficult to identify the sentiment dimension(s), especially for the two datasets composed of sentimentally dissimilar domains.

In an attempt to gain insight into why the judges found it difficult to identify the sentiment dimension(s), we show in Tables 19–22 the top-ranked features induced for each dimension using MMFR for the four augmented datasets, where the lightly shaded columns correspond to the eigenvectors chosen for the topic dimension and the darkly shaded columns correspond to the eigenvectors chosen for the sentiment dimension. After examining these results, we believe that a few points deserve mention.

First, the top features generated from the sentiment eigenvector(s) for MOV-DVD and BOO-DVD, the two datasets composed of sentimentally similar constituent domains, are clearly sentiment-oriented, making it relatively easy for the human judges to determine the sentiment eigenvector(s). Not so is the case for DVD-ELE and MOV-KIT, the two datasets composed of dissimilar domains, where the top features are "noisier" (i.e., many of them are not necessarily sentiment-oriented), thus making it tougher for the judges to locate the sentiment eigenvector(s). In fact, one can see from just the top features generated by the sentiment eigenvector(s) in Tables 19–22 that those for MOV-DVD and BOO-DVD are clearly more sentiment-oriented than those for DVD-ELE and MOV-KIT.

It should not be surprising that the more sentimentally dissimilar the constituent domains are, the noisier the top features generated from the sentiment eigenvector(s) are, however. If the constituent domains are sentimentally similar, they tend to have many sentiment-bearing words in common. This implies that these sentiment-bearing words will appear more frequently in the augmented datasets than in each of the constituent datasets. Hence, combining the two domains helps to boost the influence of these sentiment-bearing words, increasing the chance of their appearing higher up in the list of features ranked by MMFR. This reinforcement effect intuitively explains why the sentiment eigenvector is clearly dominated by sentiment words for datasets composed of sentimentally similar domains. On the other hand, if the constituent domains are sentimentally dissimilar, they tend not to have many sentiment-bearing words in common. As a result, the influence of the sentiment-bearing words that are present in only one of the two constituent domains will be "diluted" by the larger number of non-sentiment-bearing words that result from combining the two domains. In other words, the features that are clearly sentiment-oriented in just one rather than both domains may no longer appear sufficiently high in the ranked list of features. In fact, as we saw in Tables 21 and 22, the sentiment eigenvector is contaminated

| MOV-DVD | | | |
|---|---|---|---|
| $\mathbf{e}_2$ | $\mathbf{e}_3$ | $\mathbf{e}_4$ | $\mathbf{e}_5$ |
| $\mathbf{C}_1$ | $\mathbf{C}_1$ | $\mathbf{C}_1$ | $\mathbf{C}_1$ |
| roles | wonderful | recommend | kids |
| drama | excellent | fan | children |
| murder | beautiful | liked | loved |
| meets | personal | book | child |
| crime | collection | read | son |
| supporting | view | excellent | daughter |
| involving | art | amazing | boy |
| convincing | highly | definitely | school |
| tale | fantastic | highly | wonderful |
| lead | deal | absolutely | heart |
| | | | |
| $\mathbf{C}_2$ | $\mathbf{C}_2$ | $\mathbf{C}_2$ | $\mathbf{C}_2$ |
| bought | stupid | buy | quality |
| season | boring | house | dark |
| buy | dull | rent | war |
| disappointed | mean | waste | horror |
| fan | terrible | wait | release |
| amazon | save | kill | fan |
| buying | lame | murder | earth |
| copy | run | obvious | production |
| dvds | guys | season | suspense |
| watched | except | dvds | sound |

Table 19: Top ten features induced for each dimension for the MOV-DVD domain. The lightly and darkly shaded columns correspond to the topic and sentiment dimensions respectively as selected by the human judges. $\mathbf{e}_2, \ldots, \mathbf{e}_5$ are the top eigenvectors; $\mathbf{C}_1$ and $\mathbf{C}_2$ are the clusters.

| BOO-DVD | | | |
|---|---|---|---|
| $\mathbf{e}_2$ | $\mathbf{e}_3$ | $\mathbf{e}_4$ | $\mathbf{e}_5$ |
| **$\mathbf{C}_1$** | **$\mathbf{C}_1$** | **$\mathbf{C}_1$** | **$\mathbf{C}_1$** |
| reader | bought | excellent | loved |
| important | disappointed | wonderful | enjoyed |
| subject | easy | highly | children |
| understanding | information | collection | year |
| modern | price | music | wonderful |
| information | waste | special | child |
| examples | workout | classic | fun |
| political | helpful | video | son |
| business | expected | perfect | friends |
| nature | reviews | amazing | highly |
| | | | |
| **$\mathbf{C}_2$** | **$\mathbf{C}_2$** | **$\mathbf{C}_2$** | **$\mathbf{C}_2$** |
| saw | young | boring | version |
| watched | men | ending | quality |
| actors | cast | waste | waste |
| liked | role | reviews | worst |
| music | actors | couldn | review |
| season | script | novel | original |
| humor | scene | maybe | edition |
| comedy | war | pages | collection |
| favorite | performance | stupid | amazon |
| ending | action | finish | format |

Table 20: Top ten features induced for each dimension for the BOO-DVD domain. The lightly and darkly shaded columns correspond to the topic and sentiment dimensions respectively as selected by the human judges. $\mathbf{e}_2, \ldots, \mathbf{e}_5$ are the top eigenvectors; $\mathbf{C}_1$ and $\mathbf{C}_2$ are the clusters.

| DVD-ELE | | | |
|---|---|---|---|
| **e₂** | **e₃** | **e₄** | **e₅** |
| **C₁** | **C₁** | **C₁** | **C₁** |
| funny | easy | fine | video |
| acting | small | problems | card |
| family | perfect | worked | camera |
| actors | excellent | months | fast |
| action | highly | easy | easy |
| plot | nice | working | cable |
| enjoy | low | computer | picture |
| young | comfortable | day | pictures |
| wonderful | ipod | card | paper |
| comedy | headphones | drive | digital |
| | | | |
| **C₂** | **C₂** | **C₂** | **C₂** |
| unit | amazon | amazon | phone |
| battery | item | tv | waste |
| purchased | review | purchase | unit |
| device | company | disappointed | battery |
| problems | return | item | getting |
| tried | took | purchased | low |
| working | check | reviews | power |
| plug | saw | wanted | hear |
| charge | card | received | worst |
| computer | worked | ipod | batteries |

Table 21: Top ten features induced for each dimension for the DVD-ELE domain. The lightly and darkly shaded columns correspond to the topic and sentiment dimensions respectively as selected by the human judges. $\mathbf{e}_2, \ldots, \mathbf{e}_5$ are the top eigenvectors; $\mathbf{C}_1$ and $\mathbf{C}_2$ are the clusters.

| MOV-KIT | | | |
|---|---|---|---|
| $\mathbf{e}_2$ | $\mathbf{e}_3$ | $\mathbf{e}_4$ | $\mathbf{e}_5$ |
| $\mathbf{C}_1$ | $\mathbf{C}_1$ | $\mathbf{C}_1$ | $\mathbf{C}_1$ |
| james | pan | coffee | price |
| directed | cooking | clean | clean |
| sex | clean | machine | kitchen |
| hour | pans | ice | knife |
| drama | cook | maker | knives |
| relationship | heat | plastic | size |
| death | oven | cup | sharp |
| direction | heavy | fill | dishwasher |
| tv | food | months | cutting |
| michael | stick | working | attractive |
| | | | |
| $\mathbf{C}_2$ | $\mathbf{C}_2$ | $\mathbf{C}_2$ | $\mathbf{C}_2$ |
| food | months | item | pan |
| recommend | purchased | price | toaster |
| pot | worked | sheets | oven |
| purchased | broke | ordered | pans |
| mine | amazon | amazon | heat |
| kitchen | coffee | received | return |
| mixer | replacement | beautiful | bottom |
| handle | month | dishes | worked |
| size | tried | arrived | read |
| store | service | sets | toast |

Table 22: Top ten features induced for each dimension for the MOV-KIT domain. The lightly and darkly shaded columns correspond to the topic and sentiment dimensions respectively as selected by the human judges. $\mathbf{e}_2, \ldots, \mathbf{e}_5$ are the top eigenvectors; $\mathbf{C}_1$ and $\mathbf{C}_2$ are the clusters.

| | MOV-DVD | | BOO-DVD | | DVD-ELE | | MOV-KIT | |
|---|---|---|---|---|---|---|---|---|
| | Acc | ARI | Acc | ARI | Acc | ARI | Acc | ARI |
| 2nd eigenvector only | 77.1 | 0.29 | **77.8** | **0.31** | 94.2 | 0.78 | 99.3 | 0.97 |
| Top five eigenvectors | 62.4 | 0.08 | 77.2 | **0.31** | 93.9 | 0.78 | 99.3 | 0.97 |
| Interested Reader Model | **84.2** | **0.53** | 63.1 | 0.07 | **94.8** | **0.80** | **99.6** | **0.99** |
| NMF | 56.3 | 0.02 | 69.2 | 0.15 | 94.4 | 0.79 | 70.6 | 0.17 |
| Our system | 77.1 | 0.29 | **77.8** | **0.31** | 94.2 | 0.78 | 99.3 | 0.97 |

(a)

| | MOV-DVD | | BOO-DVD | | DVD-ELE | | MOV-KIT | |
|---|---|---|---|---|---|---|---|---|
| | Acc | ARI | Acc | ARI | Acc | ARI | Acc | ARI |
| 2nd eigenvector only | 54.4 | 0.01 | 52.3 | 0.01 | 50.9 | 0.00 | 50.0 | 0.00 |
| Top five eigenvectors | 68.3 | 0.13 | 52.0 | 0.00 | 50.4 | 0.00 | 50.0 | 0.00 |
| Interested Reader Model | 53.4 | 0.01 | 52.1 | 0.01 | 50.9 | 0.00 | 50.1 | 0.00 |
| NMF | 66.9 | 0.11 | 51.7 | 0.00 | 51.1 | 0.00 | **61.6** | **0.05** |
| Our system | **71.4** | **0.18** | **68.8** | **0.14** | **61.1** | **0.05** | 59.2 | 0.03 |

(b)

Table 23: Results on (a) topic-based clustering and (b) sentiment-based clustering for the four augmented datasets. The strongest results for each dataset are boldfaced.

by a number of features that are not necessarily sentiment-bearing, which make it difficult for the human judges to identify the sentiment dimension.

Another interesting point to note is that for some datasets, there seems to be more than one eigenvector that correspond to sentiment. For instance, for the BOO-DVD dataset, all five human judges agreed that both $\mathbf{e}_4$ and $\mathbf{e}_5$ correspond to the sentiment dimension. A closer examination of these two eigenvectors (shown in Table 20) reveals a very interesting pattern: in $\mathbf{e}_4$, the positive features (in $\mathbf{C}_1$) came from the DVD domain and the negative features (in $\mathbf{C}_2$) came from the BOO domain; whereas in $\mathbf{e}_5$, the positive features (in $\mathbf{C}_1$) came from BOO domain and the negative features (in $\mathbf{C}_2$) came from DVD. In other words, $\mathbf{e}_4$ partitions the reviews according to the positive of DVD and the negative of BOO, whereas $\mathbf{e}_5$ does the reverse. This suggests that the eigen-decomposition procedure is "smart" enough not to merge the positive and negative sentiment-bearing words from the two domains together. Perhaps even more importantly, both $\mathbf{e}_4$ and $\mathbf{e}_5$ are not only partitioning the reviews along the sentiment dimension but also the topic dimension.

### 4.7.2 CLUSTERING RESULTS

Rows 1–4 of Tables 23a and 23b show the topic- and sentiment-based clustering results for the same four baseline text clustering algorithms that were described in Section 4.2. Note that each of these baselines can only produce one clustering of the documents per dataset. Hence, for each baseline, the topic-based clustering results are produced by comparing this clustering against the gold-standard topic-based clustering, and the sentiment-based clustering results are produced by comparing the same clustering against the gold-standard sentiment-based clustering.

As we can see from these topic-based results in Table 23a, the baseline in which we cluster using only the second eigenvector achieves the best average clustering results over the four augmented datasets. This can potentially be attributed to the fact that $e_2$ corresponds to the topic dimension for all four datasets according to the human judges, as described in the human experiments. However, clustering using only $e_2$ does not produce the best clustering results on all four datasets. In fact, the Interested Reader Model achieves the best results on MOV-DVD, DVD-ELE, and MOV-KIT. Nevertheless, its results on BOO-DVD are the worst among the baselines. The same is true for the "top five eigenvectors" baseline and NMF: both of them have yielded poor results on MOV-DVD; in addition, NMF's results on BOO-DVD and MOV-KIT are not promising either.

As far as the sentiment-based baseline clustering results are concerned (see rows 1–4 of Table 23b), the best average performance is achieved by NMF. Except for three cases (NMF on MOV-DVD and MOV-KIT, as well as "top five eigenvectors" on MOV-DVD), these baseline results are not particularly promising, with accuracy results in the low fifties and ARI results close to zero.

The topic- and sentiment-based clustering results produced by our algorithm are shown in row 5 of Tables 23a and 23b. Specifically, these results are obtained by grouping the reviews according to the eigenvectors manually selected for the topic and sentiment dimensions, respectively. Hence, unlike the baselines, the topic-based clustering and the sentiment-based clustering produced by our algorithm are different from each other. As before, in cases where the human judges selected more than one eigenvector for each dimension, we use the eigenvector that is ranked first most frequently. As we can see, the accuracies for topic-based clustering are reasonably high, ranging from 77.1% to 99.3%. These results suggest that it is possible to achieve high-performance topic-based (or more precisely, domain-based) clustering for a dataset even when another prominent clustering dimension (i.e., sentiment) is present. On the other hand, despite the existence of eigenvectors that clearly capture the sentiment dimension for these datasets (e.g., $e_3$ for the MOV-DVD dataset), the sentiment-based clustering accuracies and ARI values are lower than those of topic-based clustering. This can potentially be attributed to the reason mentioned in the introduction: the fact that reviews are sentimentally ambiguous makes them non-trivial to classify. In comparison to the four baselines, our algorithm achieves not only the best average performance over the four datasets but also comparatively very stable performance across these datasets.

It is worth noting that the sentiment-based clustering results produced by our algorithm for MOV-DVD and BOO-DVD are higher than those for DVD-ELE and MOV-KIT. This is perhaps not surprising: as discussed before, the human judges have found it more difficult to identify the sentiment eigenvector for DVD-ELE and MOV-KIT than for MOV-DVD and BOO-DVD, owing in part to the fact that many of the top-ranked features in the sentiment eigenvector for DVD-ELE and MOV-KIT are not sentiment-oriented, which in turn can be attributed to the fact that both of these datasets correspond to domain pairs that are sentimentally dissimilar. As mentioned above, two sentimentally dissimilar constituent domains tend not to have many sentiment-bearing words in common, and consequently, the influence of the sentiment-bearing words that are present in only one of the two constituent domains will be "diluted" by the larger number of non-sentiment-bearing words that result from combining the two domains, making it difficult to produce a good sentiment-based clustering. On the other hand, combining the two domains helps to boost the influence of these sentiment-bearing words, increasing the chance of their appearing higher up in the list of features ranked by MMFR and producing a good sentiment-based clustering.

Interestingly, our algorithm achieves better topic-based clustering results on the two datasets — DVD-ELE and MOV-KIT — where it achieves poorer sentiment-based clustering results. In fact, the topic-based clustering accuracies on DVD-ELE and MOV-KIT are near perfect: 94.2% and 99.3% for DVD-ELE and MOV-KIT respectively. This is by no means a coincidence: when the constituent domains of an augmented dataset are highly dissimilar (i.e., their word usage tends to differ considerably from each other), the topic clusters are well-separated from each other and hence high topic-based clustering results can be achieved. A similar line of reasoning can explain why our algorithm finds it comparatively more difficult to produce a good topic-based clustering for MOV-DVD and BOO-DVD, where the constituent domains are similar.

These results seem to suggest that a higher topic-based accuracy/ARI implies a lower sentiment-based accuracy/ARI and vice versa. We speculate that when the constituent domains are similar, their sentiment-bearing features tend to be similar and as a result, sentiment-based results tend to be good and topic-based results tend to be poor. Additional experiments are needed to determine the reason.

Overall, these results provide supporting evidence that our feedback-oriented algorithm can produce multiple clusterings of a dataset. In particular, even though the sentiment-based clustering accuracies are not as high as the topic-based clustering accuracies for the augmented datasets, the current level of performance of our algorithm is arguably reasonable, especially considering the fact that sentiment-based clustering is a challenging task and that traditional clustering algorithms fail to even produce more than one clustering.

### 4.7.3 MULTIPLE RELEVANT EIGENVECTORS

Recall from Table 18b that for each of the four augmented datasets, there is at least one judge who indicated that more than one eigenvector is relevant to the sentiment dimension. However, when producing the sentiment-based clustering results using our system in Table 23b, we only used the eigenvector that was ranked most frequently by the human judges. To better understand whether using more relevant eigenvectors can help improve the results for sentiment-based clustering, we repeat the experiment in which we apply 2-means to cluster the documents in the space defined by all the eigenvectors that were determined as relevant by at least one judge. More specifically, we cluster with the following set of eigenvectors: {3,4} for MOV-DVD, {4,5} for BOO-DVD, {3,5} for DVD-ELE, and {3,5} for MOV-KIT.

The accuracy and ARI results of this experiment are shown in Table 24. In comparison to the results in the last row of Table 23b, we see that using additional relevant eigenvectors yields better results for all but the BOO-DVD dataset. While it may not be easy to determine the reason, we believe that the poorer results observed on BOO-DVD can be attributed to the "impurity" of $\mathbf{e}_5$, which captures not only sentiment but also topic, as discussed before. On the other hand, the additional sentiment eigenvectors chosen for the other three augmented datasets do not seem to have this impurity problem, as they all capture the sentiment dimension for only one of the constituent domains.

|            | MOV-DVD | | BOO-DVD | | DVD-ELE | | MOV-KIT | |
|------------|------|------|------|------|------|------|------|------|
|            | Acc  | ARI  | Acc  | ARI  | Acc  | ARI  | Acc  | ARI  |
| Our system | 72.2 | 0.19 | 55.7 | 0.01 | 66.2 | 0.10 | 59.8 | 0.04 |

Table 24: Results on sentiment-based clustering obtained using multiple relevant eigenvectors for the four augmented datasets.

## 5. Significance of Our Work

We believe that our approach is significant in the following aspects.

1. **Producing a clustering according to user interest.** We proposed a novel framework in which we enabled a spectral clustering algorithm to take into account human feedback and produce a clustering along the dimension of interest to the user. A particularly appealing aspect of our approach is concerned with the relatively minimal human feedback it demands, where the user just needs to take a cursory look at a small number of features that are representative of each induced dimension. It is worth noting that having a human inspect and select an automatically induced clustering dimension is a new form of interaction between a human and a clustering algorithm. It enables a human to easily engage in various clustering tasks to help improve their performance in an easy, low-effort manner. We believe that our approach, which belongs to an emerging family of interactive algorithms that allows the user to make small, guiding tweaks and thereby get results much better than would otherwise be possible, is the future of information retrieval.

2. **Inducing human-interpretable clustering dimensions.** The dimensions produced by spectral clustering or other dimensionality reduction algorithms (e.g., Latent Semantic Indexing (LSI), Deerwester, Dumais, Furnas, Landauer, & Harshman, 1990) are generally considered *non-interpretable* (Sebastiani, 2002), unlike a dimension in the original feature space, which typically corresponds to a word type and can therefore be interpreted by a human easily. The results of our preliminary study challenge this common wisdom. We show in the context of text clustering that *a dimension in the low-dimensional space induced by spectral clustering can be interpreted by a human.* We believe the ability to produce human-interpretable dimensions enables us to employ spectral clustering (and perhaps other dimensionality reduction-based clustering algorithms) for text processing in a more intelligent manner. This is especially the case with respect to selecting the dimensions that are pertinent to the task at hand. For example, in existing applications of spectral clustering to the topic-based clustering task (e.g., Xu et al., 2003; He, Cai, Liu, & Ma, 2004; Hu, Deng, Guo, & Xu, 2007), *all* of the dimensions in the low-dimensional space are typically used. Since we showed that not all dimensions produced by spectral clustering for a dataset are necessarily topic-related, we can potentially improve topic-based clustering results by *not* employing the non-topic-related dimensions in the clustering process. In addition, since some of these induced dimensions correspond to non-topic dimensions, we can use them to produce non-topic-based clusterings. In particular, given the recent surge of interest in the NLP community in text classification along non-topic dimensions such as sentiment and gender (e.g., Garera & Yarowsky, 2009; Jurafsky, Ranganath, & McFarland, 2009), our approach offers a solution to these tasks that does not rely on labeled data, unlike the majority of existing approaches to non-topic-based text classification, which are supervised in nature. Overall, we believe that NLP researchers have not fully exploited the power of spectral clustering, and hence the rewards of understanding spectral clustering in light of our results may be significant.

3. **Producing multiple clusterings.** While the majority of existing text clustering algorithms can produce a single clustering of a dataset, our approach can potentially be used to produce multiple clusterings, one along each of the important clustering dimensions induced via a novel application of spectral clustering.

Finally, it is worth mentioning that the task of inducing clustering dimensions is reminiscent of the influential *topic modeling* task (Blei, Ng, & Jordon, 2003), whose goal is to discover the major topics of a set of documents in an unsupervised manner. Note that the two tasks are fundamentally different: while a topic model attempts to discover the major topics in a set of documents, our "dimension model" aims to discover the major clustering dimensions. Nevertheless, the two models bear resemblance to each other in many ways. First, they both employ clustering to discover information from a text collection in an unsupervised manner. Second, they both display the learned information to a human using representative words: a topic model represents each induced topic using words that are representative of each topic, and our dimension model represents each induced clustering dimension using words representative of the two document clusters involved in the dimension. Finally, not all induced topics and clustering dimensions are human-recognizable, but for those that are, a human is needed to assign labels to them. We believe that the induction of clustering dimensions has the potential to substantially enhance the capability of existing text analysis algorithms to discover knowledge from a text collection in an unsupervised manner by complementing the information induced by a topic model.

## 6. Related Work

In the introduction, we discussed related work on producing a user-desired clustering. In this section, we focus on discussing related work on topic-based clustering and classification, sentiment classification, active learning, and producing multiple clusterings in computational stylistics.

**Topic-based text clustering.** Traditional research on text clustering has focused primarily on topic-based clustering, owing in large part to DARPA's Topic Detection and Tracking initiative in the 1990s. Many different clustering algorithms have been used, including non-hierarhical algorithms such as $k$-means and Expectation-Maximization (EM) and hierarchical algorithms such as single-link, complete-link, group-average, and single-pass (Hatzivassiloglou, Gravano, & Maganti, 2000). These algorithms cluster a given set of documents in a feature space that is typically spanned by all of the unigrams. However, clustering in such a high-dimensional space does not allow the distance between two documents to be reliably computed due to the *curse of dimensionality*. Consequently, more recent work has focused on the development of algorithms that cluster documents in a low-dimensional space constructed via dimensionality reduction. Representative members of this family of dimensionality reduction-based clustering algorithms include traditional algorithms that are based on LSI (Deerwester et al., 1990), as well as more recently proposed (and arguably better performing) algorithms such as spectral clustering (Shi & Malik, 2000; Ng et al., 2001), non-negative matrix factorization (Xu et al., 2003), locality preserving indexing (He et al., 2004), and locality discriminating indexing (Hu et al., 2007). Despite the development of these new clustering algorithms, they have primarily been evaluated with respect to their ability to produce topic-based clusterings.

**Topic-based text classification.** As Yang and Liu (1999) put it, text classification is inherently "a supervised learning task". In fact, it is arguably one of the most popular tasks to which supervised learning techniques were applied in the information retrieval community in the 1990s (see Sebastiani, 2002, for a comprehensive overview of related work on machine learning for text classification). Nevertheless, the annotated documents that are needed for training a high-performance supervised text classifier can be expensive to obtain. As a result, some researchers have investigated the possibility of performing text classification with little or even no labeled data. Such attempts have led to the development of general-purpose semi-supervised text classification algorithms that combine labeled and unlabeled data using transduction (Joachims, 1999b) or EM (Nigam, McCallum, Thrun, & Mitchell, 2000), the latter of which has been used in combination with active learning (McCallum & Nigam, 1998). More recently, Sandler (2005) has proposed an unsupervised text classification algorithm that is based on mixture modeling and LSI-based dimensionality reduction.

**Sentiment classification.** As mentioned in the introduction, despite the large amount of recent work on sentiment analysis and opinion mining, much of it has focused on *supervised* methods (see Pang & Lee, 2008, for a comprehensive survey of the field). One weakness of these existing supervised polarity classification systems is that they are typically *domain-* and *language-specific*. Hence, when given a new domain or language, one needs to go through the expensive process of collecting a large amount of annotated data in order to train a high-performance polarity classifier. Some recent attempts have been made to leverage existing sentiment corpora or lexicons to automatically create annotated resources for new domains or languages. However, such methods require the existence of either a parallel corpus/machine translation engine for projecting/translating annotations/lexicons from a resource-rich language to the target language (Banea, Mihalcea, Wiebe, & Hassan, 2008; Wan, 2008), or a domain that is "similar" enough to the target domain (Blitzer et al., 2007). When the target domain or language fails to meet this requirement, sentiment-based clustering and unsupervised polarity classification become appealing alternatives. Unfortunately, with a few exceptions (e.g., semi-supervised sentiment analysis, Riloff & Wiebe, 2003; Sindhwani & Melville, 2008; Dasgupta & Ng, 2009a; Li, Zhang, & Sindhwani, 2009), these tasks are largely under-investigated in the NLP community. Turney's (2002) work is perhaps one of the most notable examples of unsupervised polarity classification. However, while his system learns the semantic orientation of the phrases in a review in an unsupervised manner, this information is used to predict the polarity of a review heuristically.

**Domain adaptation.**  Domain adaptation, also known as transfer learning, has been one of the focal research areas in machine learning and NLP in recent years, where the goal is to leverage the labeled data available for one domain (the source domain) to build a classifier for another domain (the target domain). Techniques for domain adaptation has been applied to various NLP tasks, including part-of-speech tagging, noun phrase chunking, syntactic parsing, named entity recognition, and word sense disambiguation (e.g., Daumé III & Marcu, 2006; Chan & Ng, 2007; Duamé III, 2007; Jiang & Zhai, 2007a, 2007b). Of particular relevance to our work are domain adaptation techniques specifically developed for text and sentiment classification (e.g., Blitzer, McDonald, & Pereira, 2006; Finn & Kushmerick, 2006; Blitzer et al., 2007; Gao, Fan, Jiang, & Han, 2008; Ling, Dai, Xue, Yang, & Yu, 2008; Tan, Cheng, Wang, & Xu, 2009). It is worth noting that our domain adaptation setting is different from the traditional setting. Traditionally, sophisticated classifiers and/or an automatically constructed mapping of features between the two domains are used in the adaptation process. In our setting, however, we simply utilize the sentiment dimension that is manually selected for the source domain to automatically identify the sentiment dimension for the target domain.

**Active clustering.**  Active learning is a heavily investigated machine learning paradigm that aims to achieve better generalization bounds with lower annotation costs (Cohn, Atlas, & Ladner, 1994). While in a traditional active learning setting, a human is requested to annotate the *data points* that a classifier is most uncertain about (e.g., Cohn et al., 1994), recent research in active learning has involved asking a human to identify or label the features that are useful for the classification task at hand (e.g., Bekkerman et al., 2007; Raghavan & Allan, 2007; Druck, Settles, & McCallum, 2009; Roth & Small, 2009). As mentioned in the introduction, active learning has been applied in a clustering setting, with the goal of "encouraging" an algorithm to produce the user-intended clustering when the data can be clustered along multiple dimensions. Different variants of active clustering have been proposed. Some request a human to label a pair of data points as "must-link" or "cannot-link" to indicate whether the two points must or must not reside in the same cluster (e.g., Wagstaff et al., 2001; Bilenko, Basu, & Mooney, 2004), while others have a human determine whether two clusters should be merged or split during a hierarchical clustering process (e.g., Balcan & Blum, 2008). Our active clustering algorithm is yet another variant: we ask a human to select the clustering she desires from a set of automatically produced clusterings.

**Generation of multiple clusterings.**  The notion that text collections may be clustered in multiple independent ways has been discussed in the literature on computational stylistics (see Lim, Lee, & Kim, 2005; Biber & Kurjian, 2006; Grieve-Smith, 2006; Tambouratzis & Vassiliou, 2007; Gries, Wulff, & Davies, 2010, for example). In machine learning, there have been attempts to design algorithms for producing multiple clusterings of a dataset. While some of them operate in a semi-supervised setting (e.g., Gondek & Hofmann, 2004; Davidson & Qi, 2007), some are totally unsupervised (e.g., Caruana, Elhawary, Nguyen, & Smith, 2006; Jain, Meka, & Dhillon, 2008). For instance, Caruana et al.'s (2006) meta clustering algorithm produces $m$ different clusterings of a dataset by running $k$-means $m$ times, each time with a random selection of seeds and a random weighting of features. Its goal is to present each local minimum found by $k$-means as a possible clustering. However, they do not propose any mechanism for determining which of these $m$ clusterings is the one the user desires. Our approach, which relies on spectral clustering rather than $k$-means for producing multiple clusterings, fills this gap by soliciting user feedback to determine the user-desired clustering.

## 7. Conclusions and Future Work

Unsupervised clustering algorithms typically group objects along the most prominent dimension, in part owing to their objective of simultaneously maximizing inter-cluster similarity and intra-cluster dissimilarity. Hence, if the user's intended clustering dimension is not the most prominent dimension, these unsupervised clustering algorithms will fail miserably. To address this problem, we proposed an active clustering algorithm, which allows us to mine the user-intended, possibly hidden, dimension of the data and produce the desired clustering. This mechanism differs from competing methods in that it requires very limited feedback: to select the intended dimension, the user only needs to inspect a small number of features. We demonstrated its viability via a set of human and automatic experiments with the challenging, yet under-investigated task of sentiment-based clustering, obtaining promising results. Additional experiments provided suggestive evidence that (1) domain adaptation can be successfully applied to identify the sentiment dimension for a new domain if the domains under consideration are sentimentally similar; (2) a hand-crafted subjectivity lexicon, if available, can be used to replace the user feedback needed to select the sentiment eigenvector of a domain; and (3) our algorithm can potentially be used to produce multiple clusterings for datasets that possess multiple clustering dimensions.

Equally importantly, we empirically demonstrated that it is possible for a human to interpret a dimension produced by a spectral clustering algorithm, contrary to the common wisdom that the dimensions in an automatically constructed rank-reduced space are non-interpretable. We believe that NLP researchers have not fully exploited the power of spectral clustering, and hence the rewards of understanding spectral clustering in light of our results may be significant. Finally, our proposal to represent an induced clustering dimension as sets of informative features facilitates exploratory text analysis, potentially enhancing the capability of existing text analysis algorithms by complementing the information provided by other unsupervised models (e.g., a topic model).

In future work, we plan to explore several extensions to our active clustering algorithm. First, as our active clustering algorithm can potentially be used to produce multiple clusterings of a dataset, one interesting future direction would be to examine its theoretical guarantees, determining whether it is able to produce distinct clusterings that are qualitatively strong (see Dasgupta & Ng, 2010a, 2010b, for example). Second, we plan to use our algorithm in combination with existing feedback-oriented methods (e.g., Bekkerman et al., 2007; Roth & Small, 2009) for improving its performance. For instance, instead of having the user construct a relevant feature space from scratch, she can simply extend the set of informative features identified for the user-selected dimension. Third, since none of the steps in our algorithm is specifically designed for sentiment classification, we plan to apply it to other non-topic-based text classification tasks that have recently received a lot of in-terest in the NLP community, such as gender classification (i.e., the task of determining the gender of the author of a document). Finally, we plan to adopt a richer representation of a document that exploits features such as polarity-oriented words obtained from hand-built or machine-learned sentiment lexicons (e.g., Hu & Liu, 2004; Wiebe, Wilson, Bruce, Bell, & Martin, 2004; Andreevskaia & Bergler, 2006; Mohammad, Dunne, & Dorr, 2009; Rao & Ravichandran, 2009), or those derived from finer-grained (i.e., sentential, sub-sentential, phrase-based) sentiment analysis methods (e.g., Wilson et al., 2005; Kennedy & Inkpen, 2006; Polanyi & Zaenen, 2006; McDonald, Hannan, Neylon, Wells, & Reynar, 2007; Choi & Cardie, 2008), as richer features may make it further easier for the user to identify the desired dimension when using our method.

## Bibliographic Note

Portions of this work were previously presented in a conference publication (Dasgupta & Ng, 2009b). The current article extends this work in several ways, most notably: (1) a detailed introduction to spectral clustering (Section 2.2); (2) the inclusion of two more baseline systems (Section 4.2); (3) an investigation of the effect of document representation on clustering performance (Section 4.4); (4) the addition of three new sections focusing on issues in domain adaptation (Section 4.5), employing a manually constructed subjectivity lexicon (Section 4.6), and producing multiple clusterings of a dataset (Section 4.7); as well as (5) a description of the significance of our work (Section 5).

## Acknowledgments

The authors acknowledge the support of National Science Foundation (NSF) grant IIS-0812261. We thank the four anonymous reviewers for their helpful comments and for unanimously recommending this article for publication in JAIR. Any opinions, findings, conclusions or recommendations expressed in this article are those of the authors and do not necessarily reflect the views or official policies, either expressed or implied, of NSF.